\documentclass{PoS}

\usepackage{my-defs}
\usepackage{wrapfig}
\usepackage{caption}
\usepackage{xspace}
\usepackage{cite}
\usepackage{lineno}
\graphicspath{figs/}

\title{Electroweak probes with ATLAS}

\ShortTitle{Electroweak probes with ATLAS}

\author{\speaker{Alexander Milov} for the ATLAS Collaboration \\
        Weizmann Institute of Science\\
        E-mail: \email{alexander.milov@weizmann.ac.il}}

\abstract{Measuring electroweak bosons in relativistic heavy ion collisions at high energy provide an opportunity to understand temporal evolution of the quark-gluon plasma created in such collisions by constraining the initial state of the interaction. Due to lack of color charges the bosons and or particles produced in their leptonic decays are unaffected by the quark-gluon plasma and therefore preserve the information about the very early stage of the collision when they were born. This singles electroweak bosons as a unique and very interesting class of observables in heavy ion collisions.

The ATLAS experiment at LHC measures production of electroweak bosons in \pp, \pPb, and \PbPb collisions systems. A review of the existing results is given in this proceeding that includes studies made with isolated photons to constraint kinematic properties and flavour composition of associated jets, measurements of  $W$ and $Z$ bosons used to estimate nuclear modification of parton distribution function and the production rates of the bosons used to verify geometric models implied to estimate event centrality. A novel analysis on measuring two-particle correlations in \pp collisions where the $Z$ boson is registered is also discussed in the proceeding.  This is the first attempt to break into the initial geometry of the \pp collisions by constraining the impact parameter with a hard scattering process. It shows that the strength of the two particle correlations in such collision is $1.08\pm0.06$ above the inclusive. To make the measurement ATLAS solves the technical problem of measuring the underlying event in high pileup condition.}

\FullConference{
12th International Workshop on High-pT Physics in the RHIC/LHC Era\\
        2-5 October, 2017\\
        University of Bergen, Bergen, Norway}

\begin{document}

\section{Introduction}
The heavy ion research program at the LHC opened wide opportunity to study electroweak (EW) probes in HI collisions. These probes carry unique information about the initial stage of HI interactions. The ATLAS experiment~\cite{ATLAS} performs measurements with high energy isolated photons, $Z$, and $W$ bosons. 

\section{Energy balance}
\begin{figure}[htb!]
\begin{center}
\begin{minipage}{.38\textwidth}
\includegraphics[width=\textwidth]{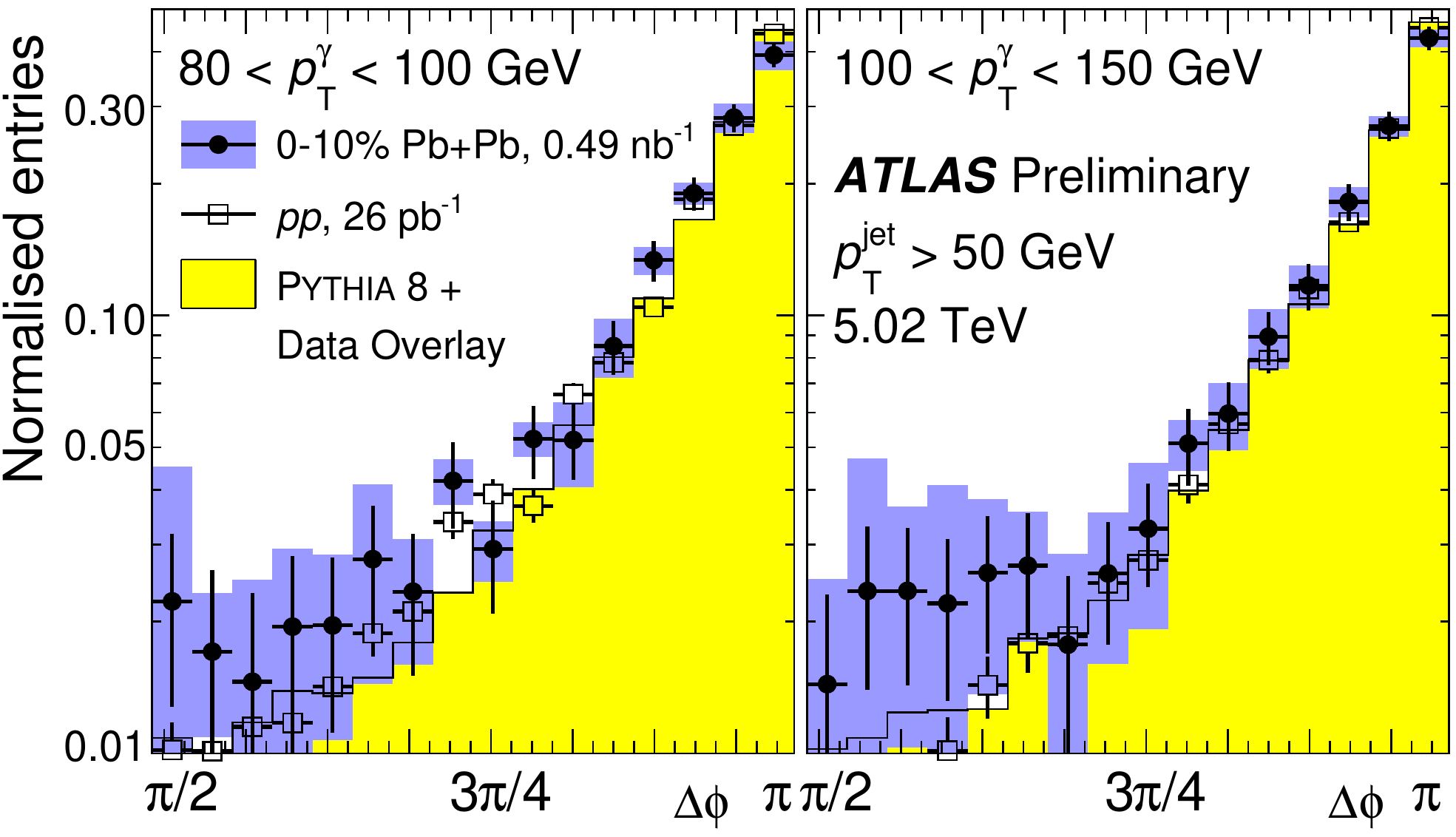}
\end{minipage}
\hspace{3mm}
\begin{minipage}{.22\textwidth}
\includegraphics[width=\textwidth]{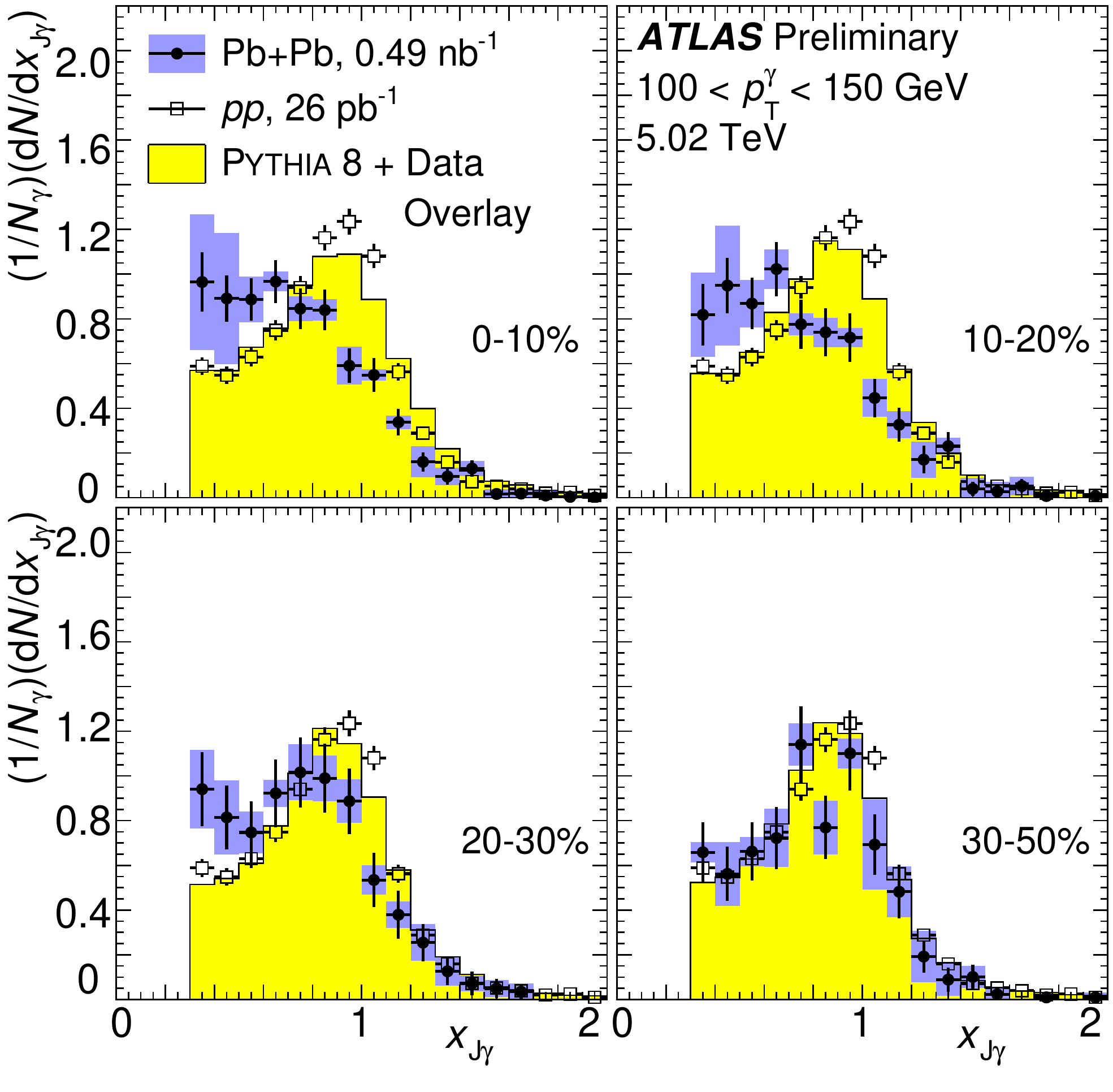} 
\end{minipage}
\begin{minipage}{.35\textwidth}
\includegraphics[width=\textwidth]{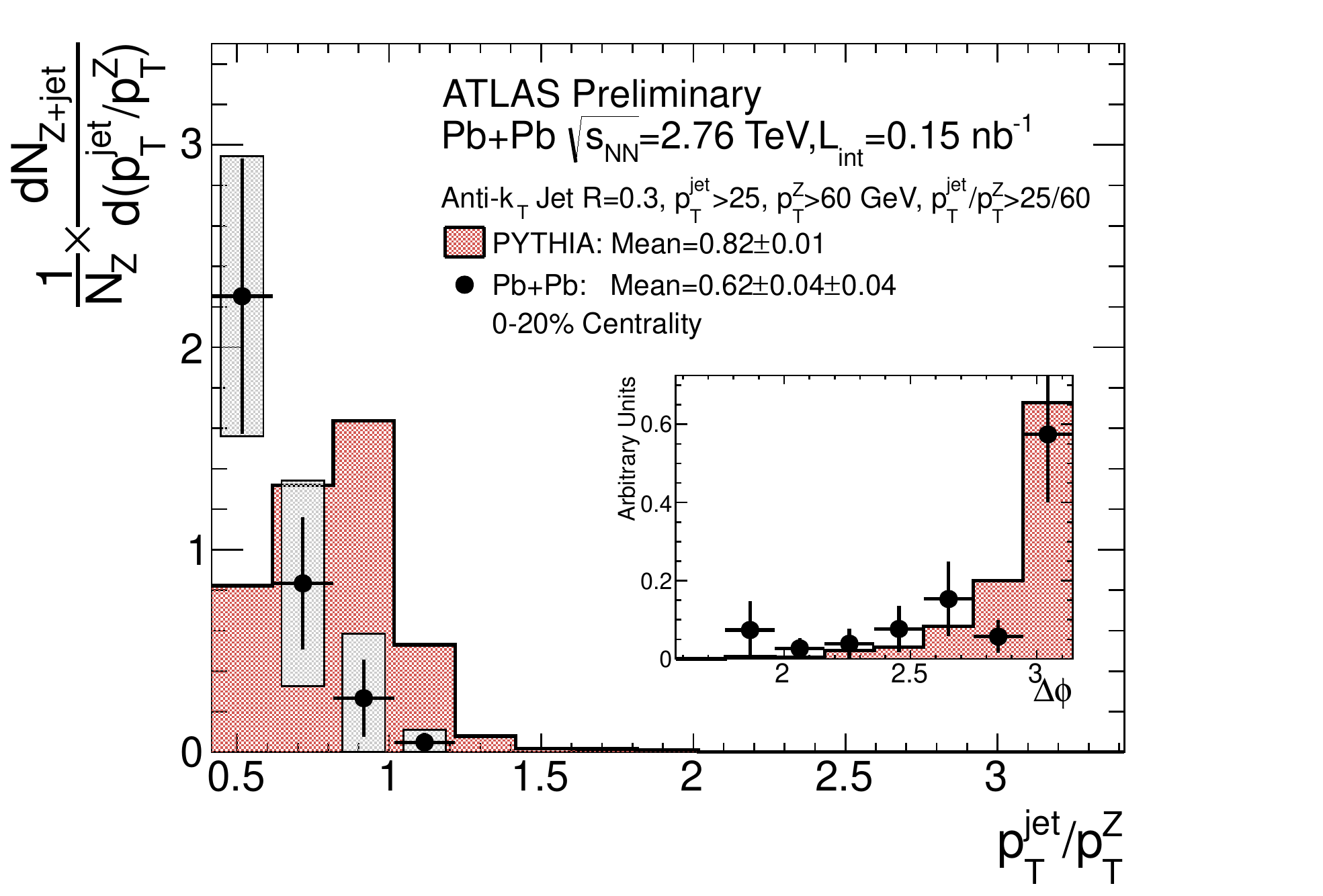}
\end{minipage}
\end{center}
\goup
\caption{Left panels: $\gamma$--jet angular correlations in central \PbPb collisions. Middle panels $\gamma$--jet energy balance in different centralities. Right: $Z$--jet energy balance in \PbPb~\cite{ATLAS-CONF-2016-110}. \label{fig:slide4}}
\end{figure}

Accurate determination of the photon or $Z$-boson momentum allows constraining the energy of the jet produced in the same scattering. As shown in Fig.~\ref{fig:slide4} 
energetic jets measured in \PbPb collisions at $\sqn=5.02$~TeV are found to be produced almost back-to-back with the boson at all centralities. At the same time, their energy balance expressed by the ratio $X_{J\gamma}$ strongly depends on centrality as it is also shown in the figure. In central collisions $X_{J\gamma}$ distribution loses its characteristic peak seen in peripheral and in the \pp collisions, and the mean of the $X_{J\gamma}$ distribution diminishes. Analogous measurement can also be performed by ATLAS with larger statistical samples using $Z$ boson--jet pairs. The proof of principle measurement at 2.76 TeV is also shown in the figure.

\begin{figure}[htb!]
\begin{center}
\begin{minipage}{.30\textwidth}
\vspace{-2mm}
\includegraphics[width=0.9\textwidth]{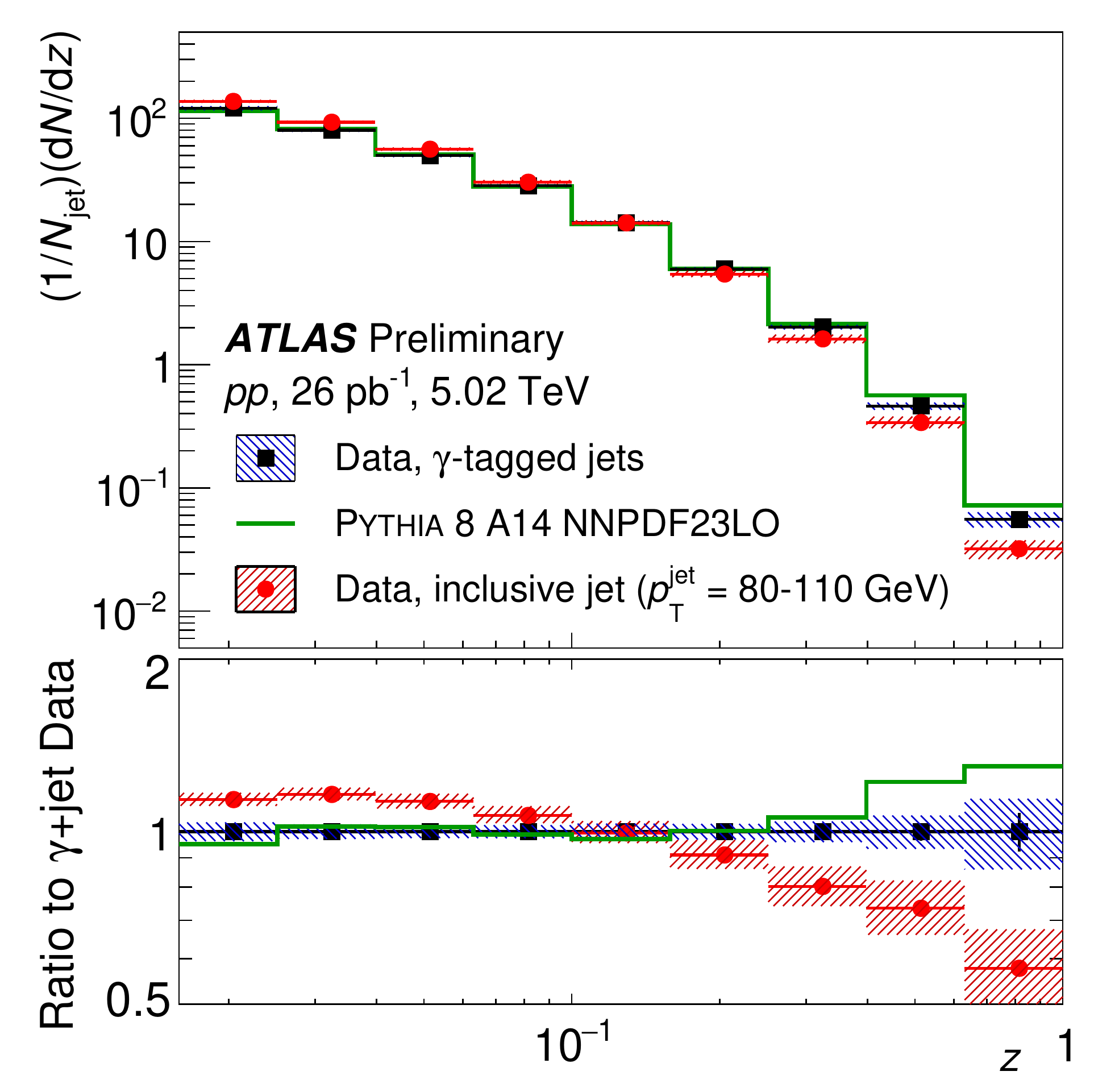}
\end{minipage}
\begin{minipage}{.33\textwidth}
\includegraphics[width=0.9\textwidth]{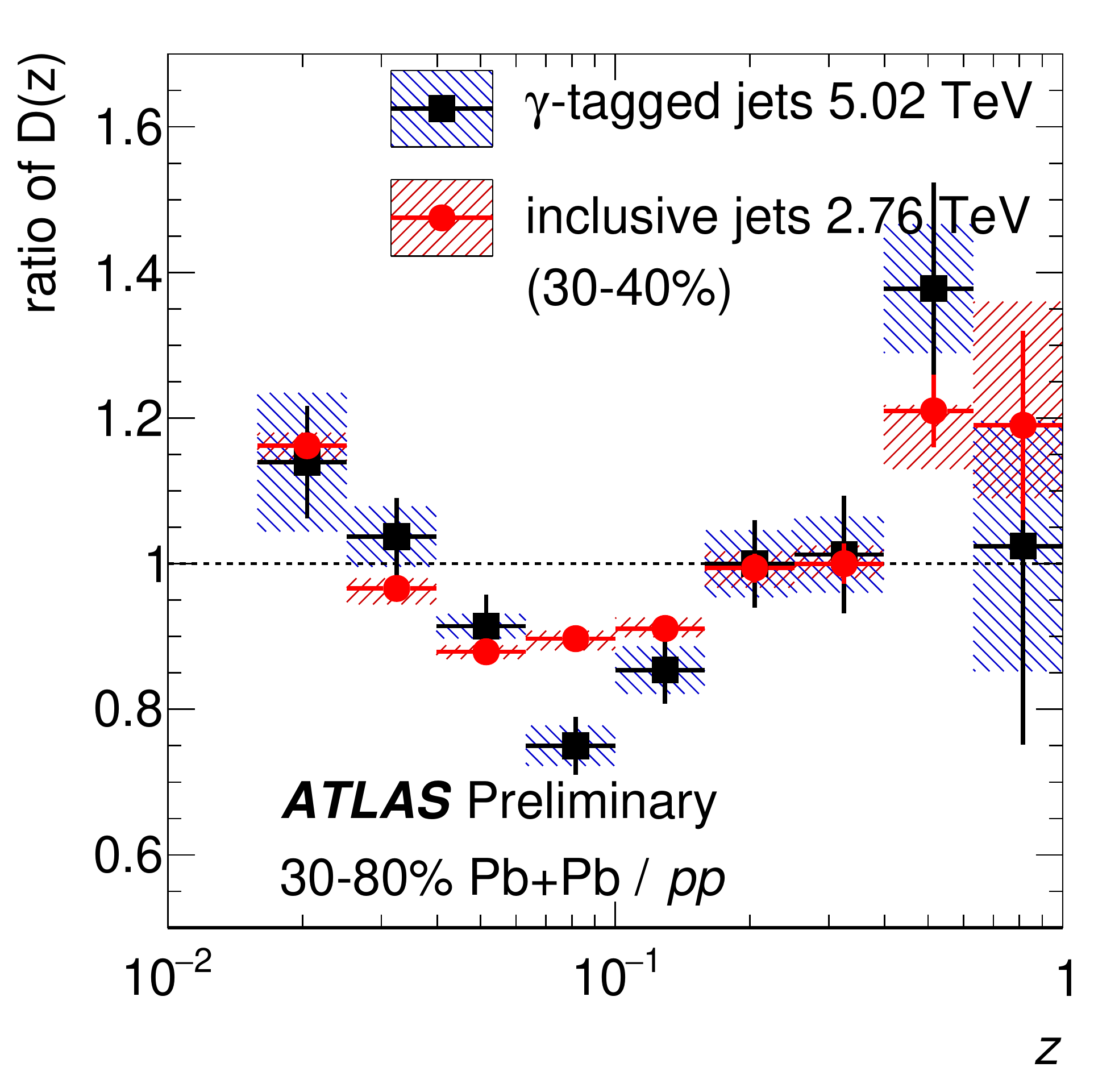}
\end{minipage}
\begin{minipage}{.33\textwidth}
\includegraphics[width=0.9\textwidth]{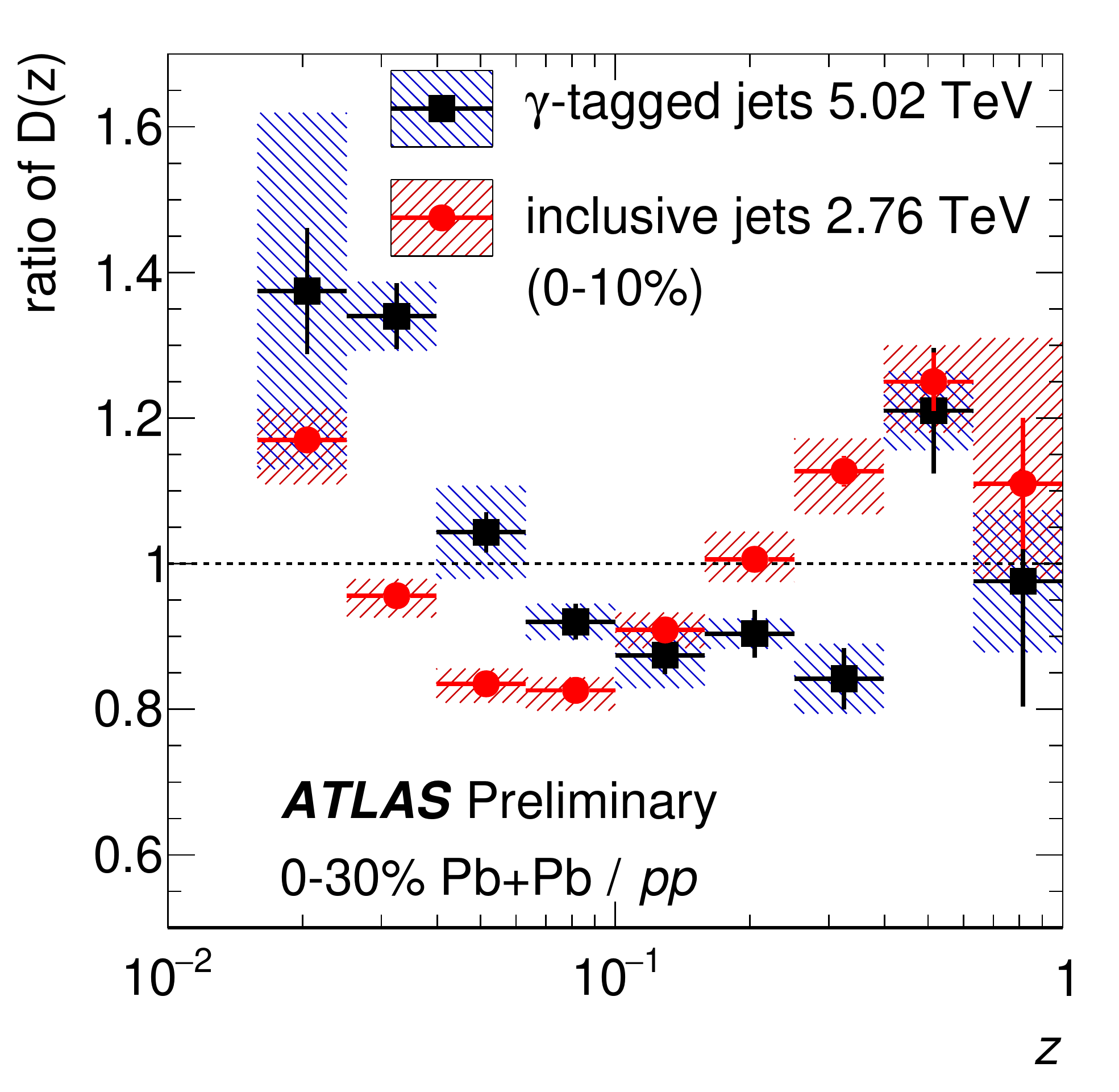}
\end{minipage}
\end{center}
\goup
\caption{Left: fragmentation function measured for $\gamma$--tagged and inclusive jets. Middle and right panels: Ratio of fragmentation function measured in mid-central and central \PbPb to that in peripheral for $\gamma$--tagged and inclusive jets~\cite{ATLAS-CONF-2017-074}. \label{fig:slide5}}
\end{figure}

Selecting jet produced in pairs with photons allows to tag jet predominantly produced in quark scattering. Figure~\ref{fig:slide5} shows 
comparison of fragmentation functions of jets associated with photons and the inclusive jets measured in \PbPb and in \pp. In more central collisions shown in the right panel, such jets are modified stronger than in the inclusive collisions at the same centrality.

\section{Parton distribution function}
\begin{figure}[htb!]
\begin{center}
\begin{minipage}{.30\textwidth}
\includegraphics[width=\textwidth]{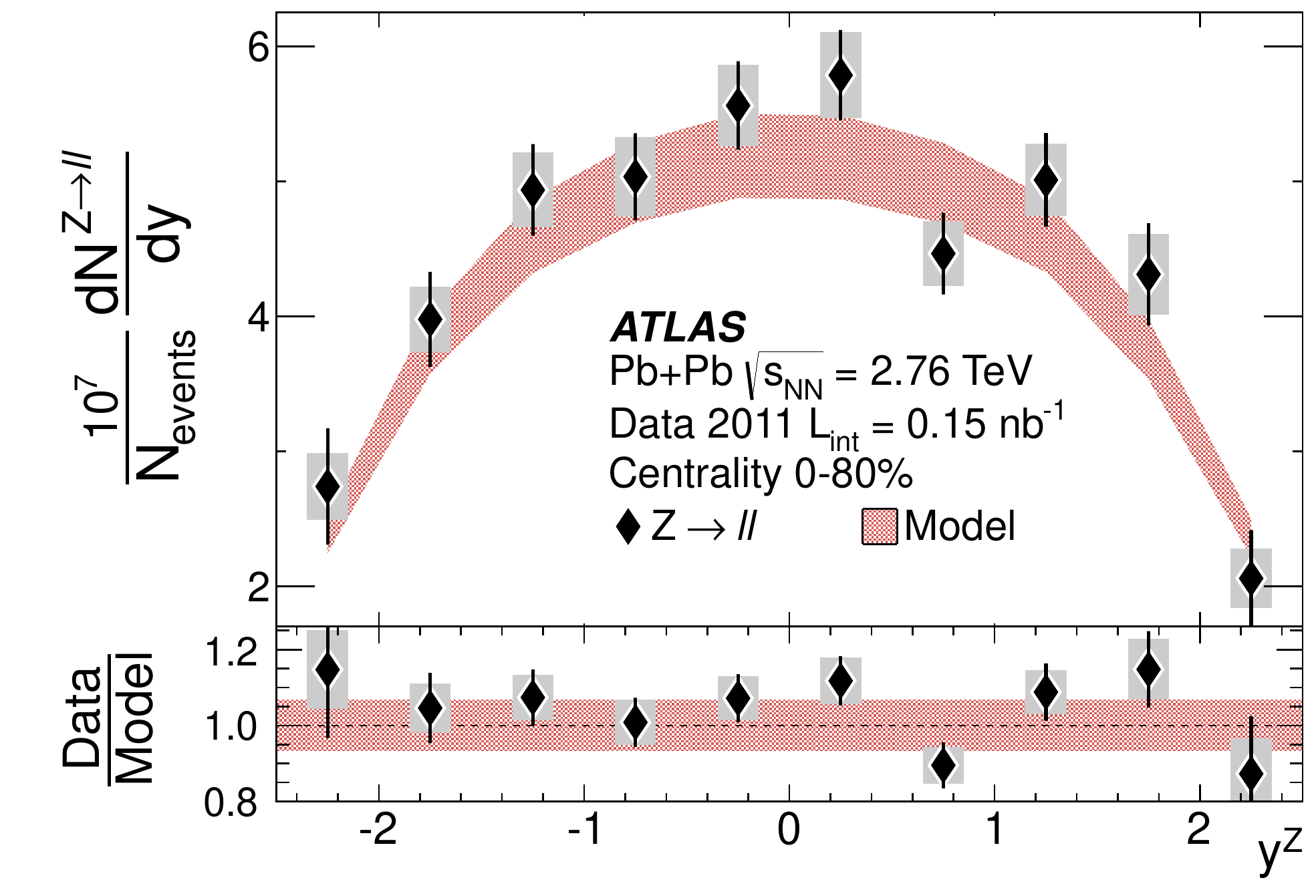} 
.\,\,\includegraphics[width=0.96\textwidth]{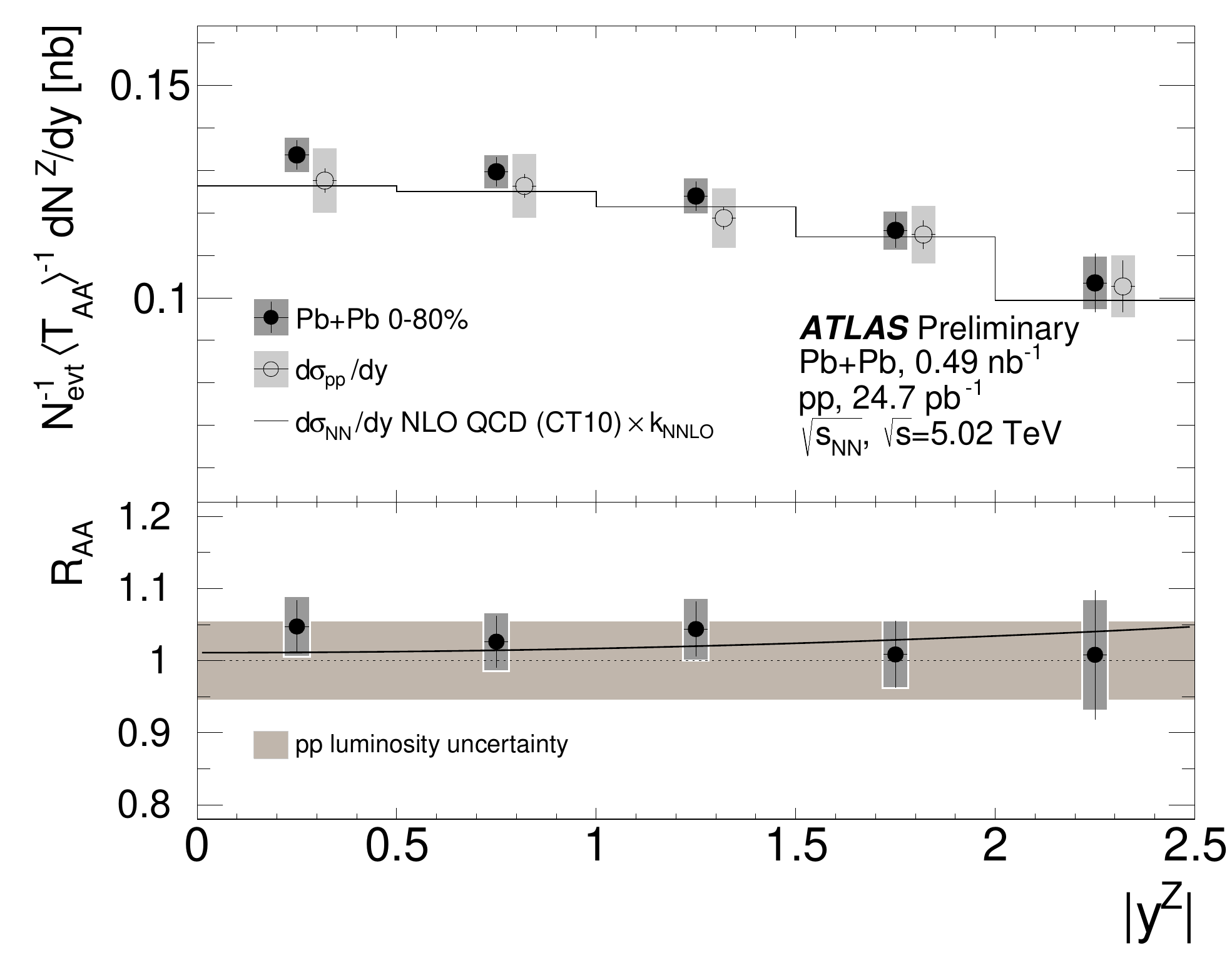}
\end{minipage}
\begin{minipage}{.33\textwidth}
\includegraphics[width=\textwidth]{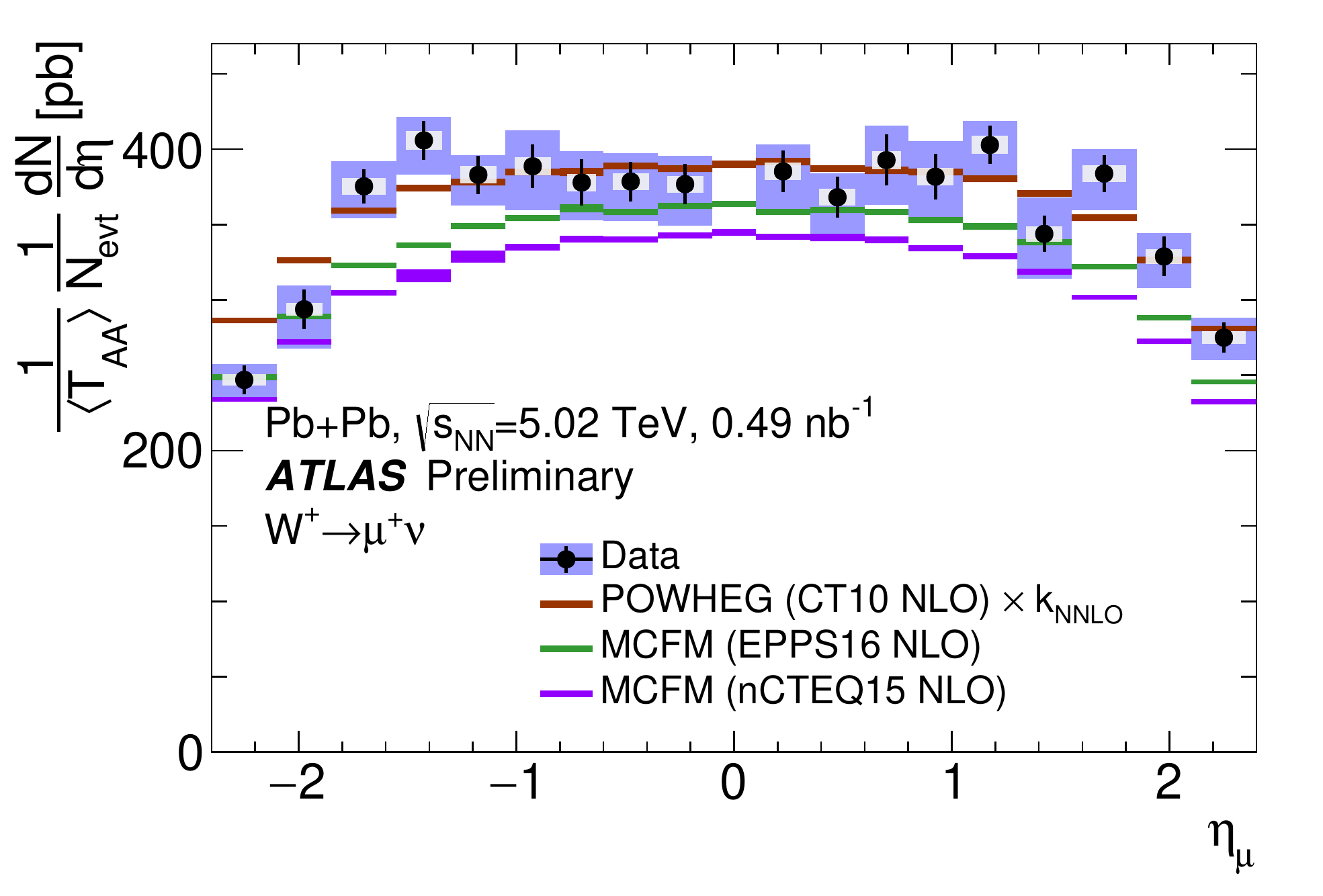}
\includegraphics[width=\textwidth]{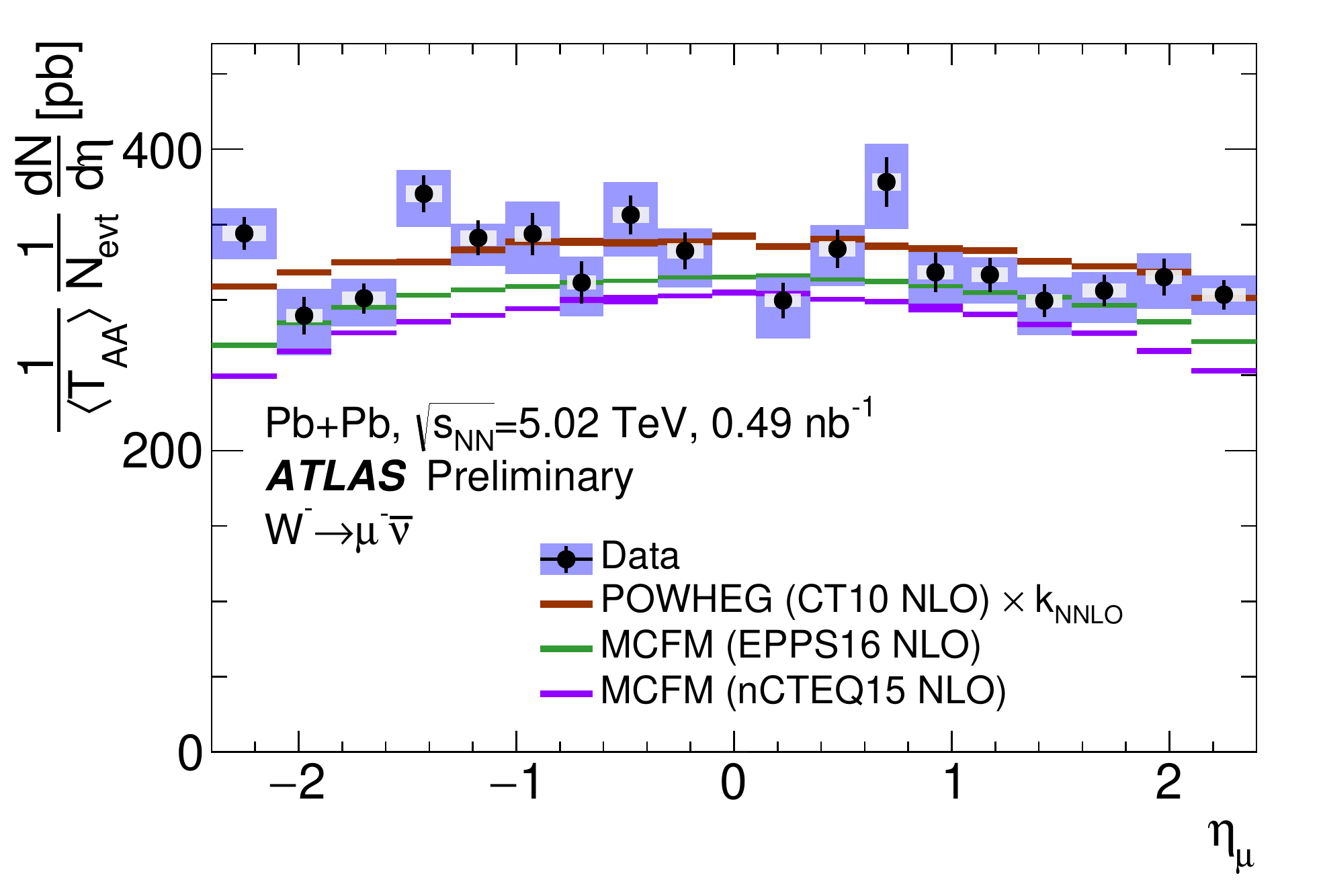}
\end{minipage}
\begin{minipage}{.27\textwidth}
\includegraphics[width=\textwidth]{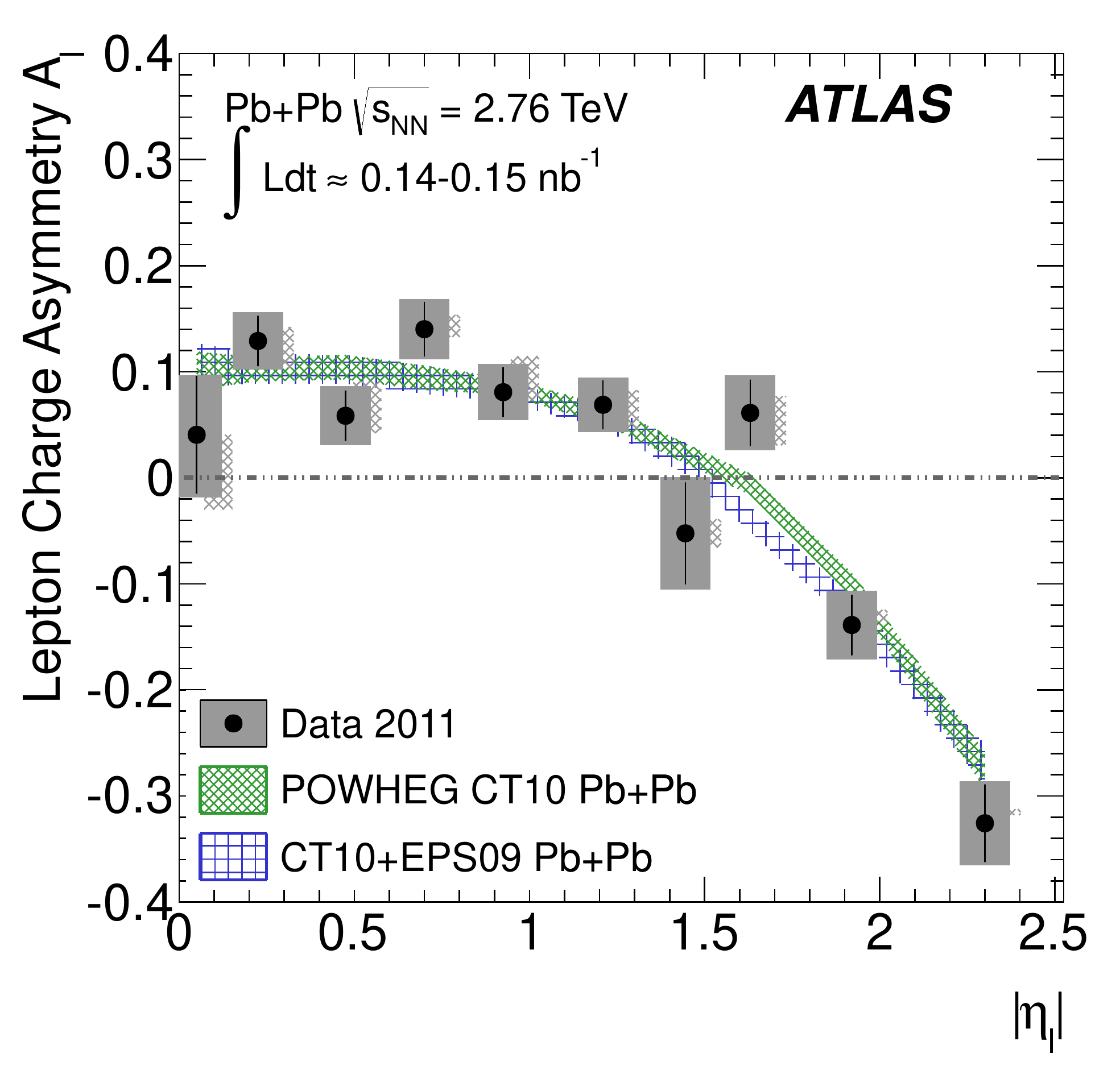}
\includegraphics[width=\textwidth]{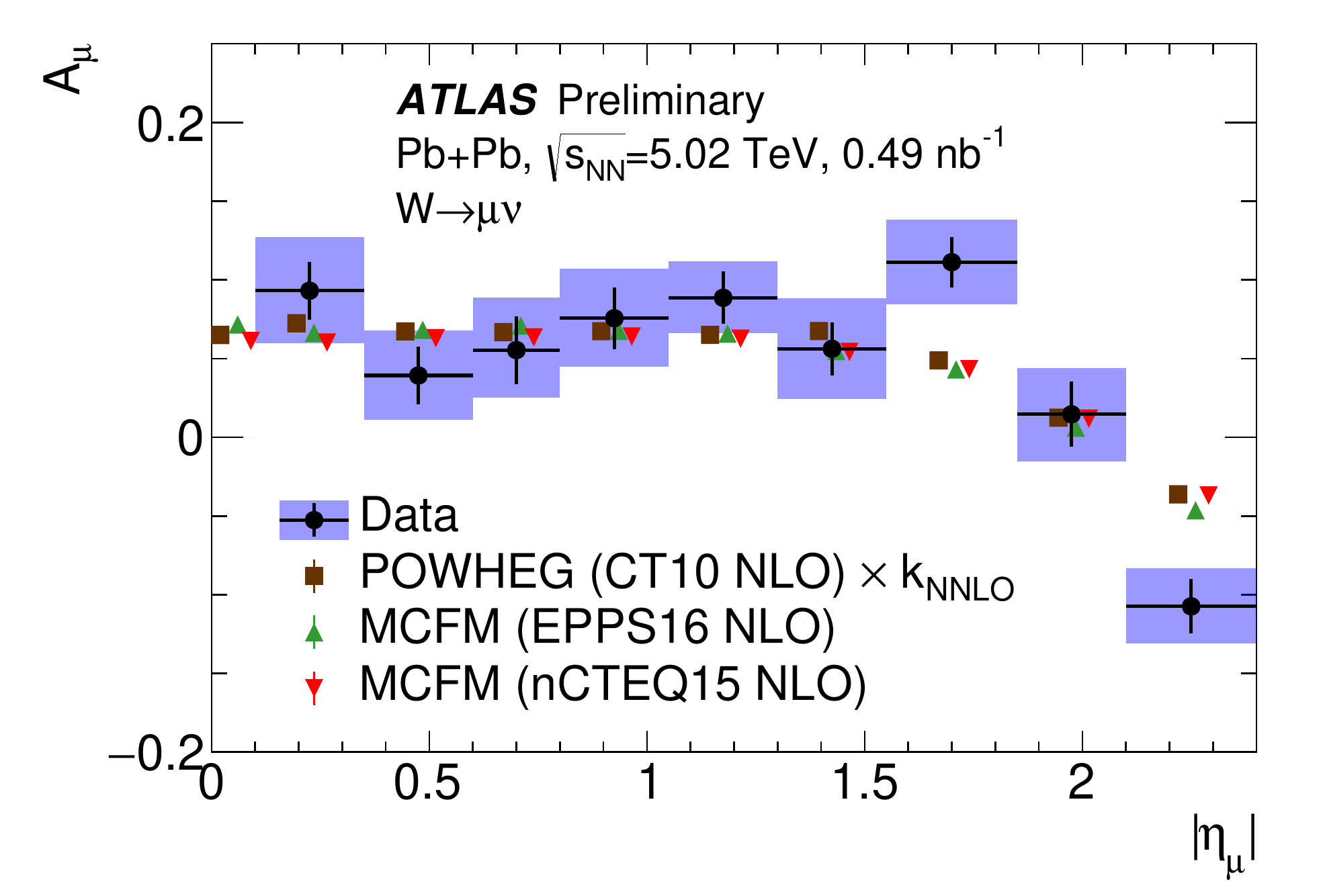}
\end{minipage}
\end{center}
\goup
\caption{Left panels: rapidity distributions for $Z$ boson measured in \PbPb and their ratio to \pp model or data. Middle panels: pseudorapidity  distribution of muons coming from $W^{\pm}$ decays in \PbPb and compared to model calculations. 
Right panels: $W$ boson charge asymmetry~\cite{PRL110-022301, ATLAS-CONF-2017-010}.
\label{fig:slide8-10}}
\end{figure}

A unique possibility to reconstruct kinematics of EW bosons unaffected by the medium created in HI interactions makes them a perfect observable to measure nuclear modification of parton distribution functions (nPDF). Measurements performed with $Z$ boson at two energies are shown in Fig.~\ref{fig:slide8-10} 
It shows that the NLO calculation based on CT10 PDF~\cite{CT10} and corrected for the NNLO cross section well reproduces the \PbPb data. However, the nPDF modifications using EPS09 and EPPS16 or nCTEQ15~\cite{npdf}, are also consistent with the data. Higher statistics is required to make an accurate measurement of nPDF modification. The $W$ lepton charge asymmetry is shown in the right panels. Although it allows some systematic uncertainty cancellation, in practice is not more sensitive to nPDFs than the rapidity distributions.

\begin{figure}[htb!]
\begin{center}
\begin{minipage}{.55\textwidth}
\includegraphics[width=\textwidth]{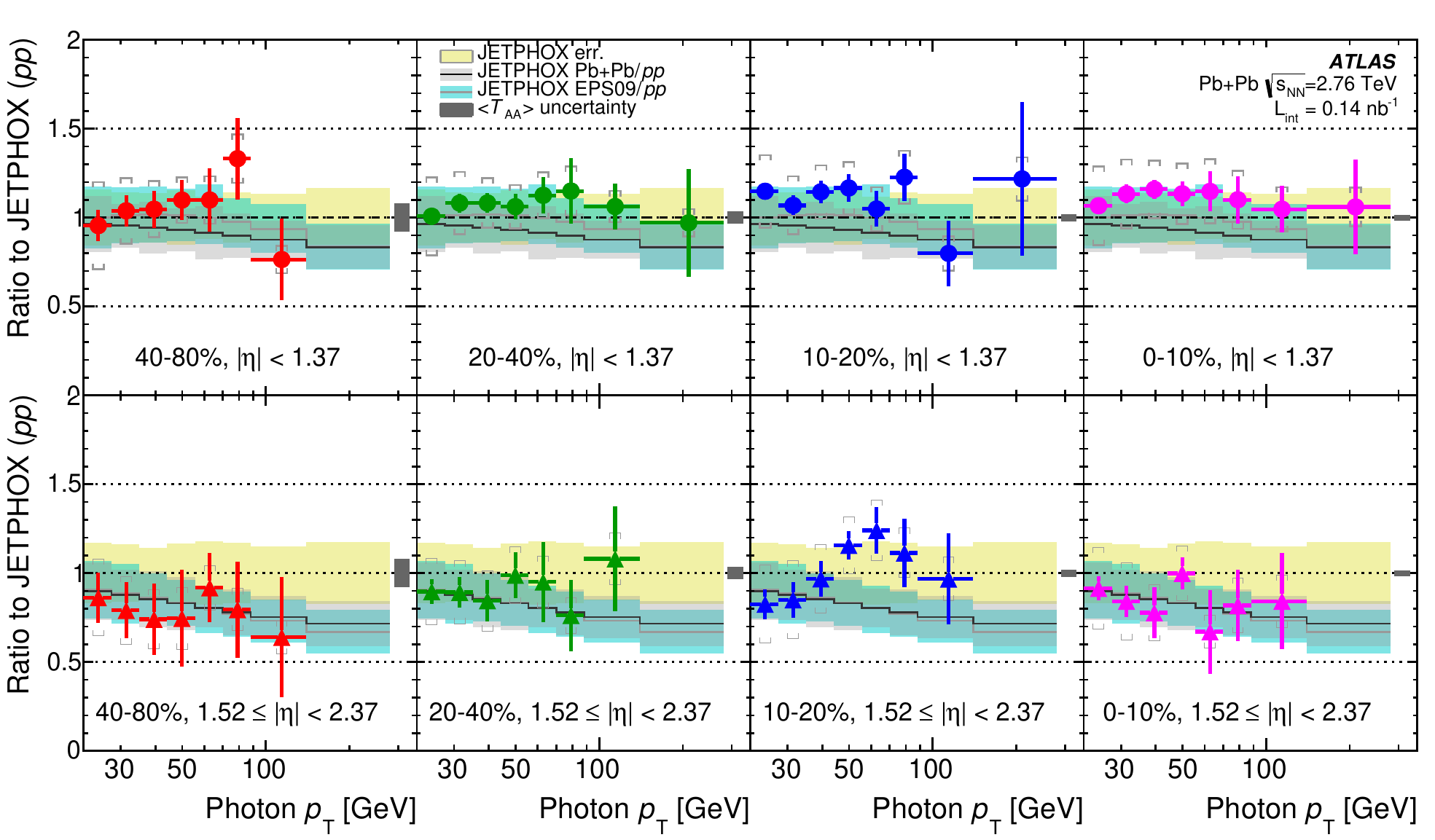}
\end{minipage}
\begin{minipage}{.44\textwidth}
\includegraphics[width=\textwidth]{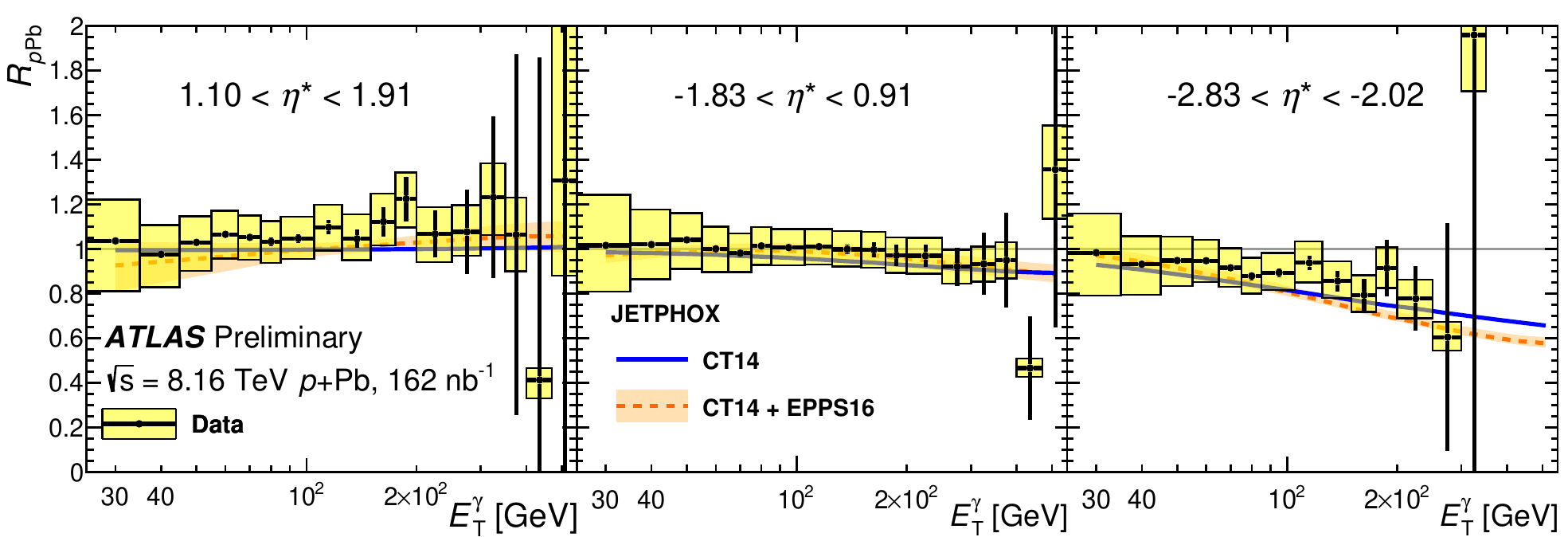}
\includegraphics[width=\textwidth]{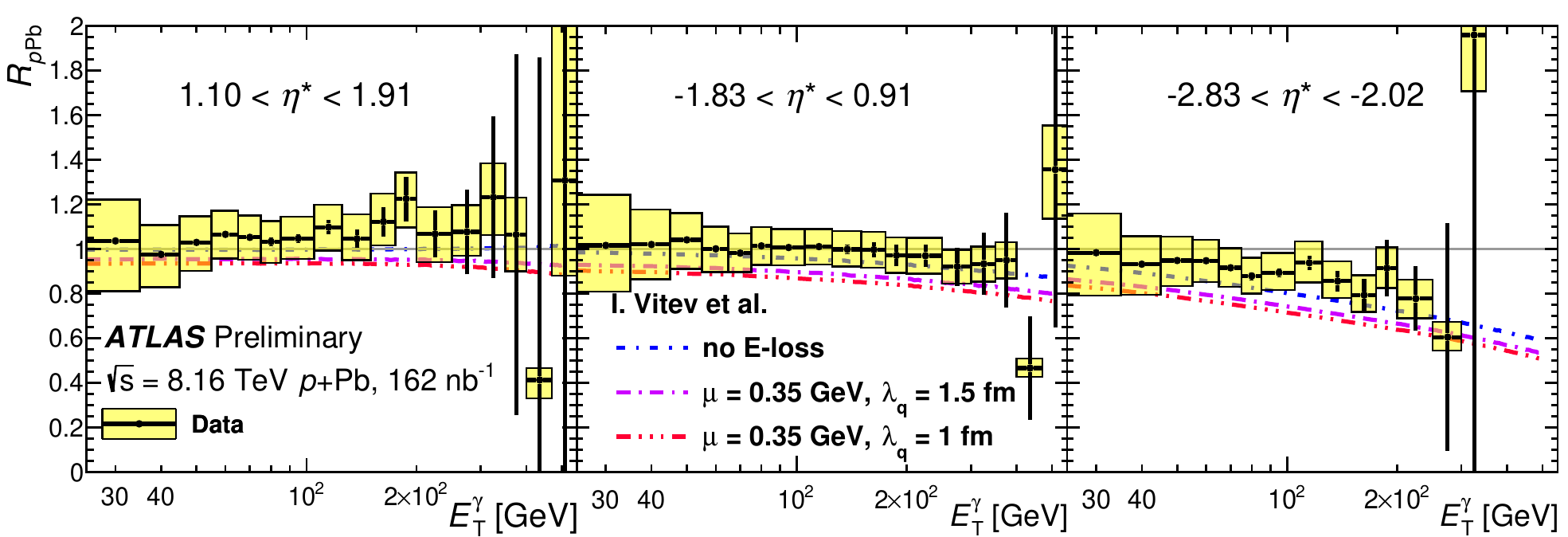}
\end{minipage}
\end{center}
\goup
\caption{Left panels: ratios of the photon spectra to model prediction for \PbPb at different rapidities and centralities. Right panels: photon \RpPb compared to model predictions~\cite{PRC93-034914,ATLAS-CONF-2017-072}.
\label{fig:slide7-12}}
\end{figure}

Results on the measurements of isolated high-\pT\ photons in \PbPb at $\sqn=2.76$~TeV made by ATLAS are compared to model predictions using nPDF as shown in left panels of Fig.~\ref{fig:slide7-12}. 
The scaled \pp predictions are found to agree well with the data in all centrality and $\eta$ regions, within the stated statistical and systematic uncertainties. Data give slight preference to the calculation using EPS09 PDF set. This situation repeats itself also for the \pPb data at $\sqn=8.16$~TeV, shown in the right panels, although at high rapidity the modelled dependencies run below the data. This is also true for \pp data.

\begin{figure}[htb!]
\begin{center}
\begin{minipage}{.24\textwidth}
\includegraphics[width=\textwidth]{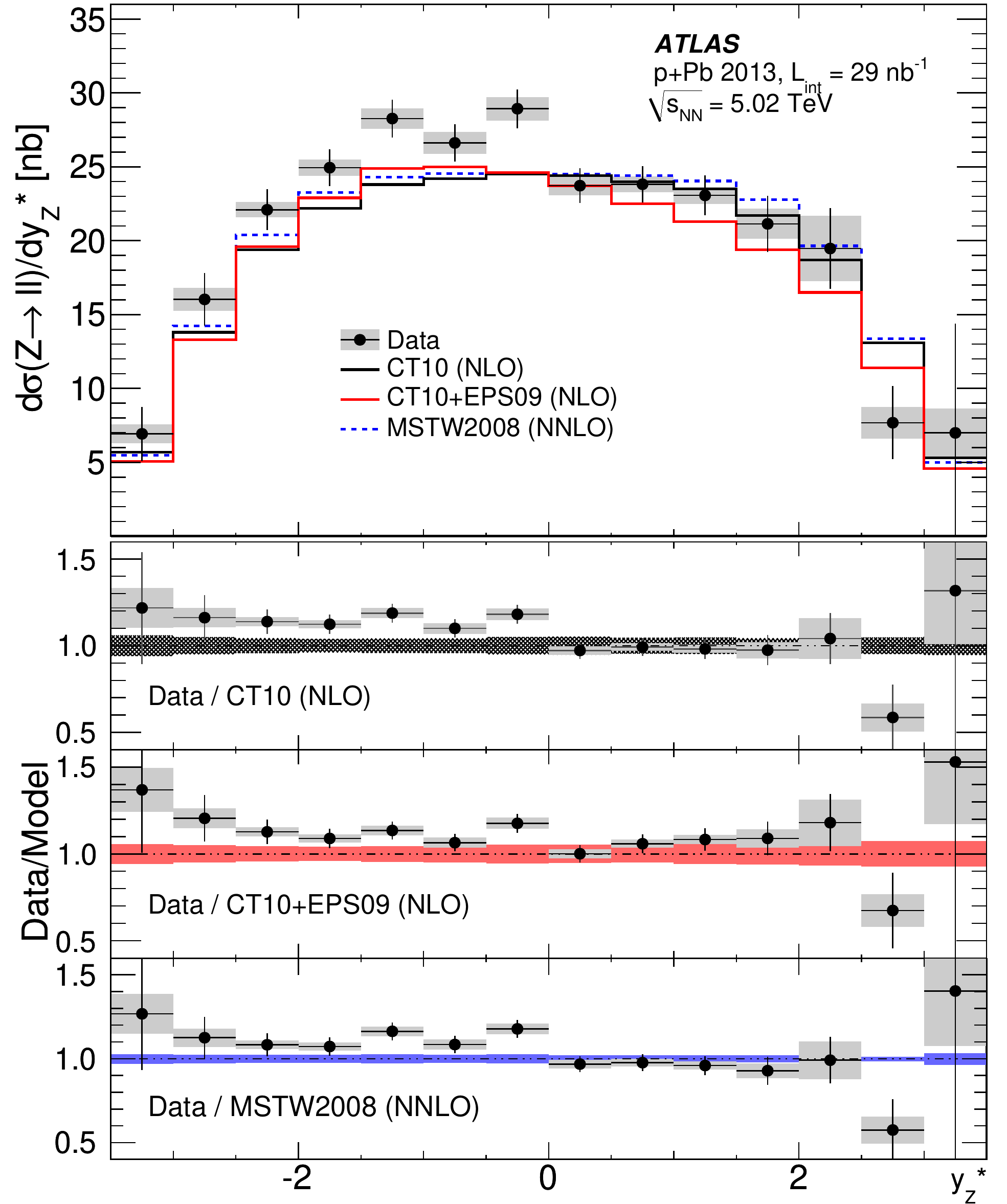}
\end{minipage}\hspace{3mm}
\begin{minipage}{.25\textwidth}
\includegraphics[width=\textwidth]{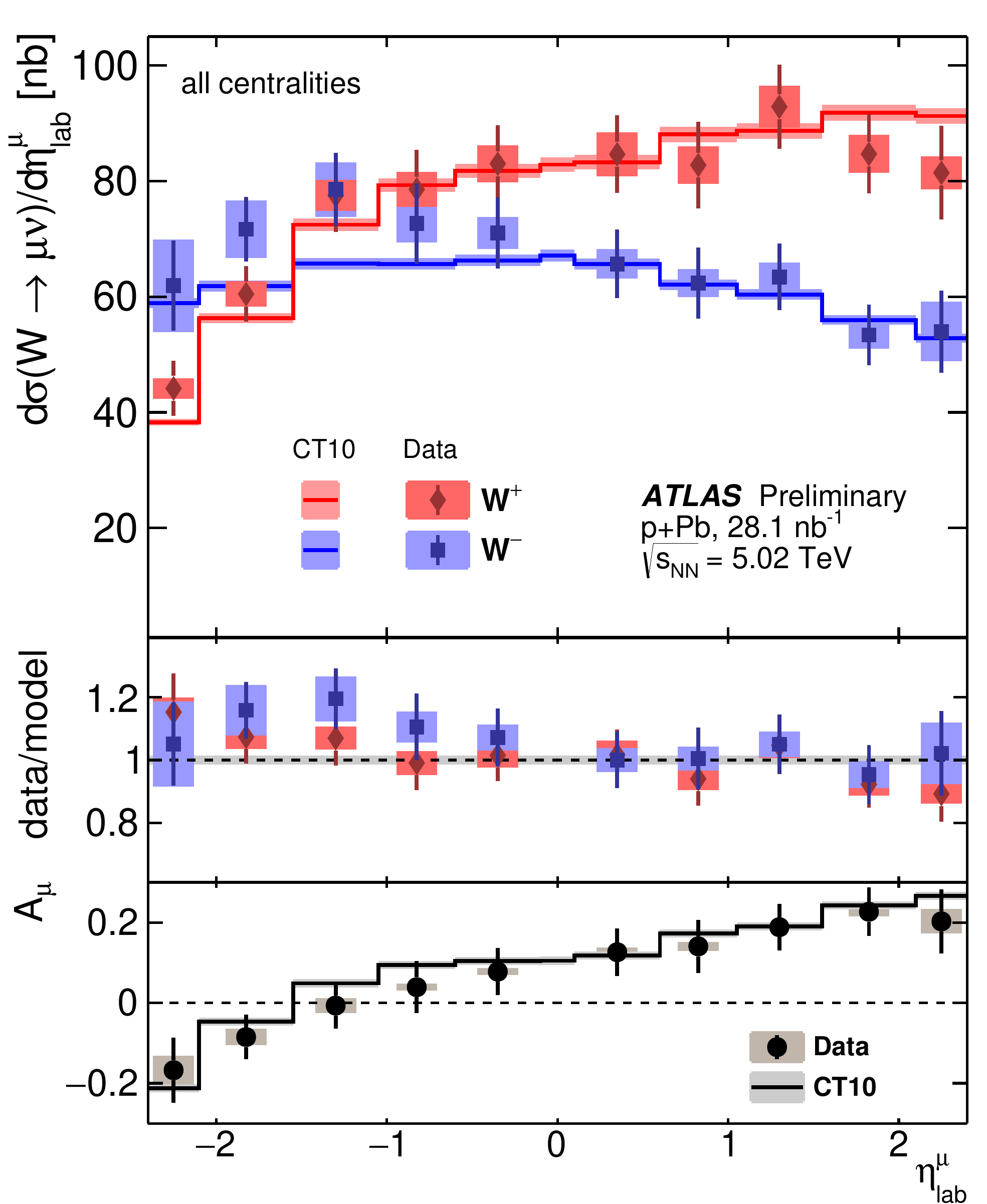}
\end{minipage}
\begin{minipage}{.45\textwidth}
\includegraphics[width=\textwidth]{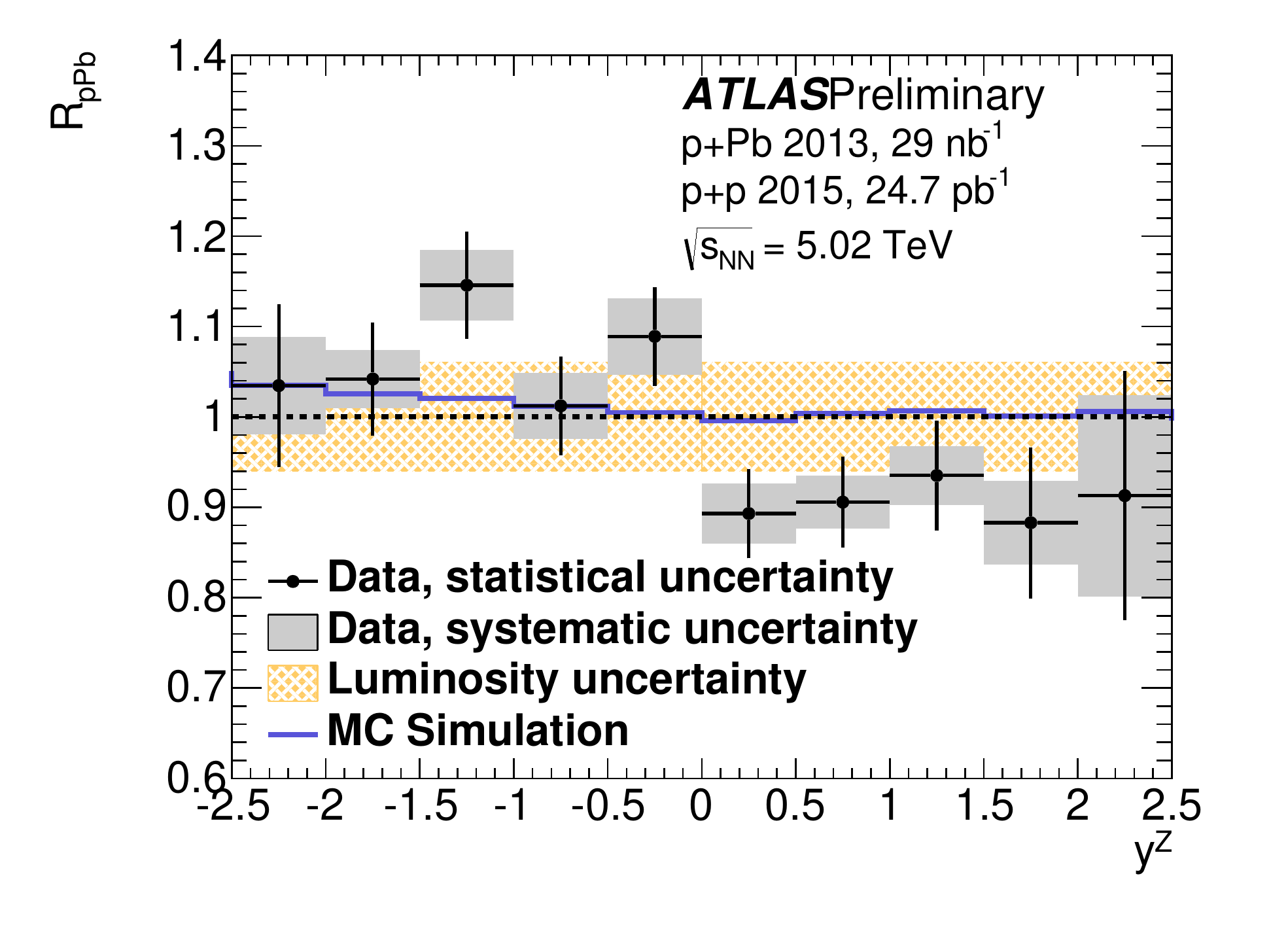}
\end{minipage}
\end{center}
\goup
\caption{Left panels: rapidity distributions for $Z$ boson measured in \pPb compared to models. Middle panels: pseudorapidity  distribution of muons coming from $W^{\pm}$ measured in \pPb decays compared to model calculations. 
Right panels: $Z$ boson \RpPb~\cite{PRC92-044915,ATLAS-CONF-2015-056,ATLAS-CONF-2016-107}.
\label{fig:slide13}}
\end{figure}

In addition to photons, the ATLAS experiment makes a comprehensive study of massive the EW boson production in \pPb collisions at $\sqn=5.02$~TeV. The results are presented in Fig.~\ref{fig:slide13}. 
The comparison of the data for $Z$ and $W$ bosons to the model calculations using the CT10 and EPS09 sets are shown in the left panels for the two particles respectively. Overall the agreement is good, with a slight preference to the EPS09 set, however, there is a clear trend in the data that it is higher than the model predictions on the left (Pb-fragmenting) side of distributions. This is true for both measured species. The effect is also somewhat larger in central \pPb collisions, whereas in peripheral the data are consistent with the calculations. The preliminary result on the $Z$-boson \RpPb shown in the right panel of the figure suggests that the misbalance seen in the ratios with models is rather a deficit on the $p$-fragmentation side than an excess on the Pb-fragmentation side. A more detailed study is needed to get an answer to this question.

\section{Initial state geometry in \NucNuc}
\begin{figure}[htb!]
\begin{center}
\begin{minipage}{.24\textwidth}
\includegraphics[width=0.98\textwidth]{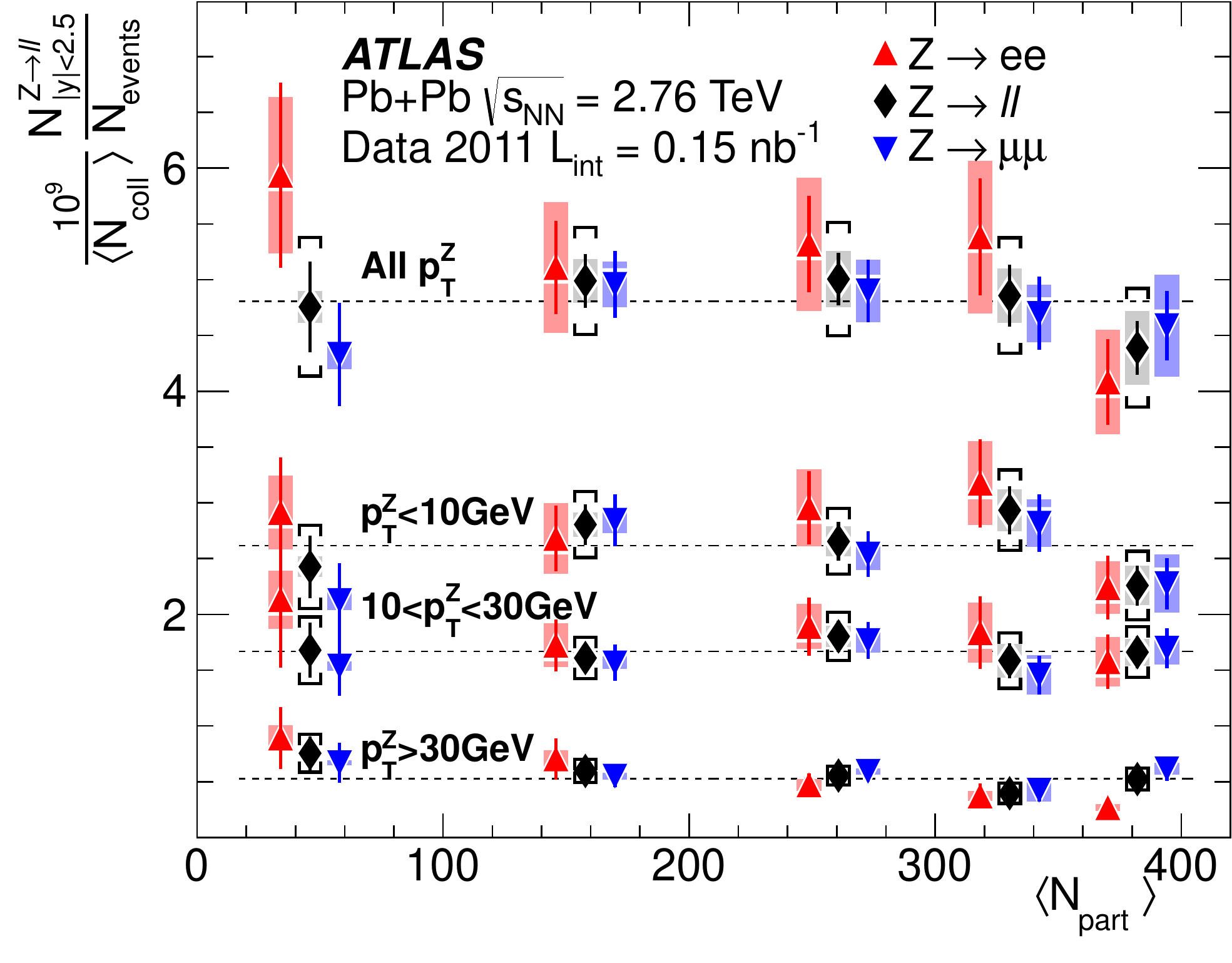}
\end{minipage}
\begin{minipage}{.24\textwidth}
\includegraphics[width=\textwidth]{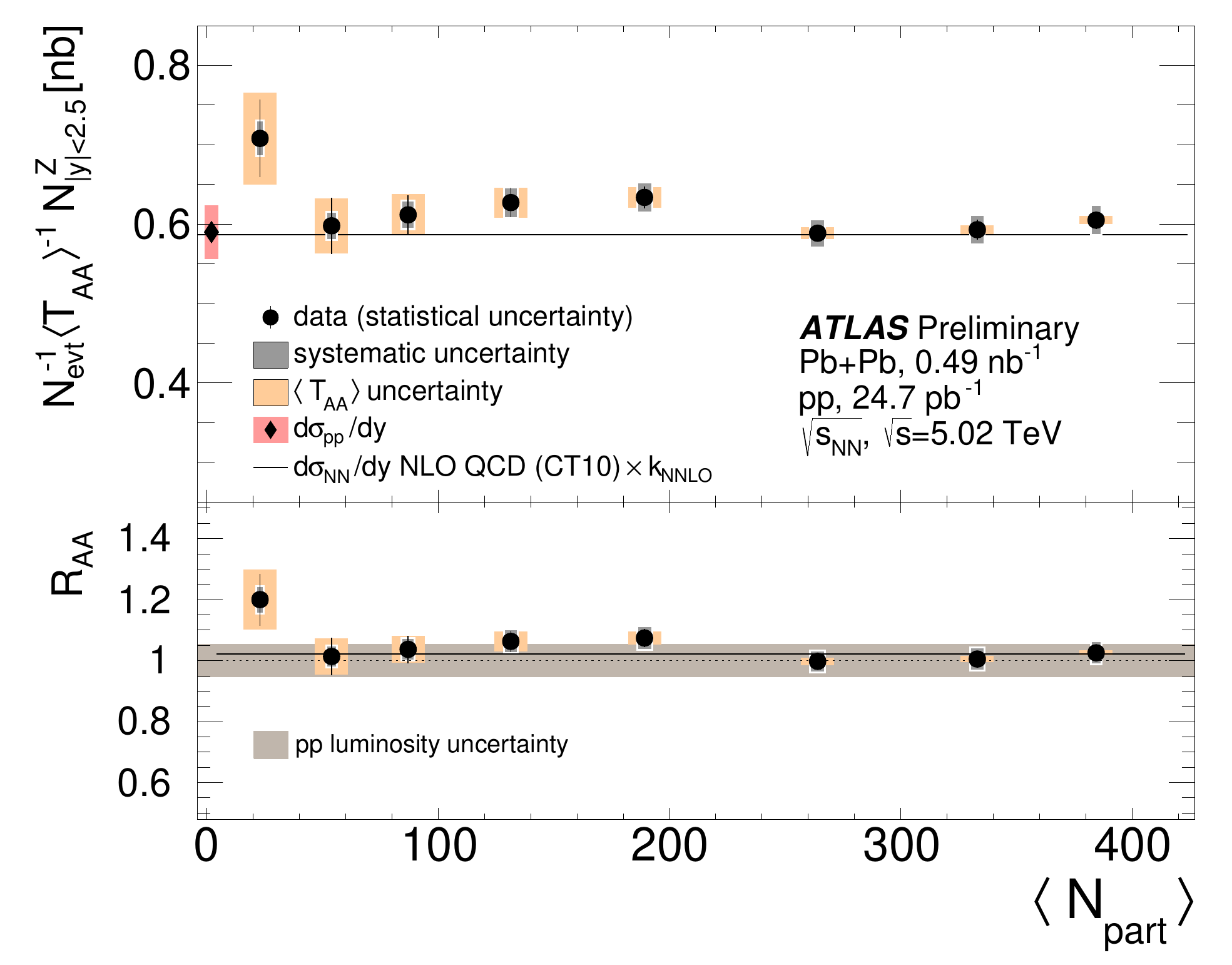}
\end{minipage}
\begin{minipage}{.21\textwidth}
\includegraphics[width=\textwidth]{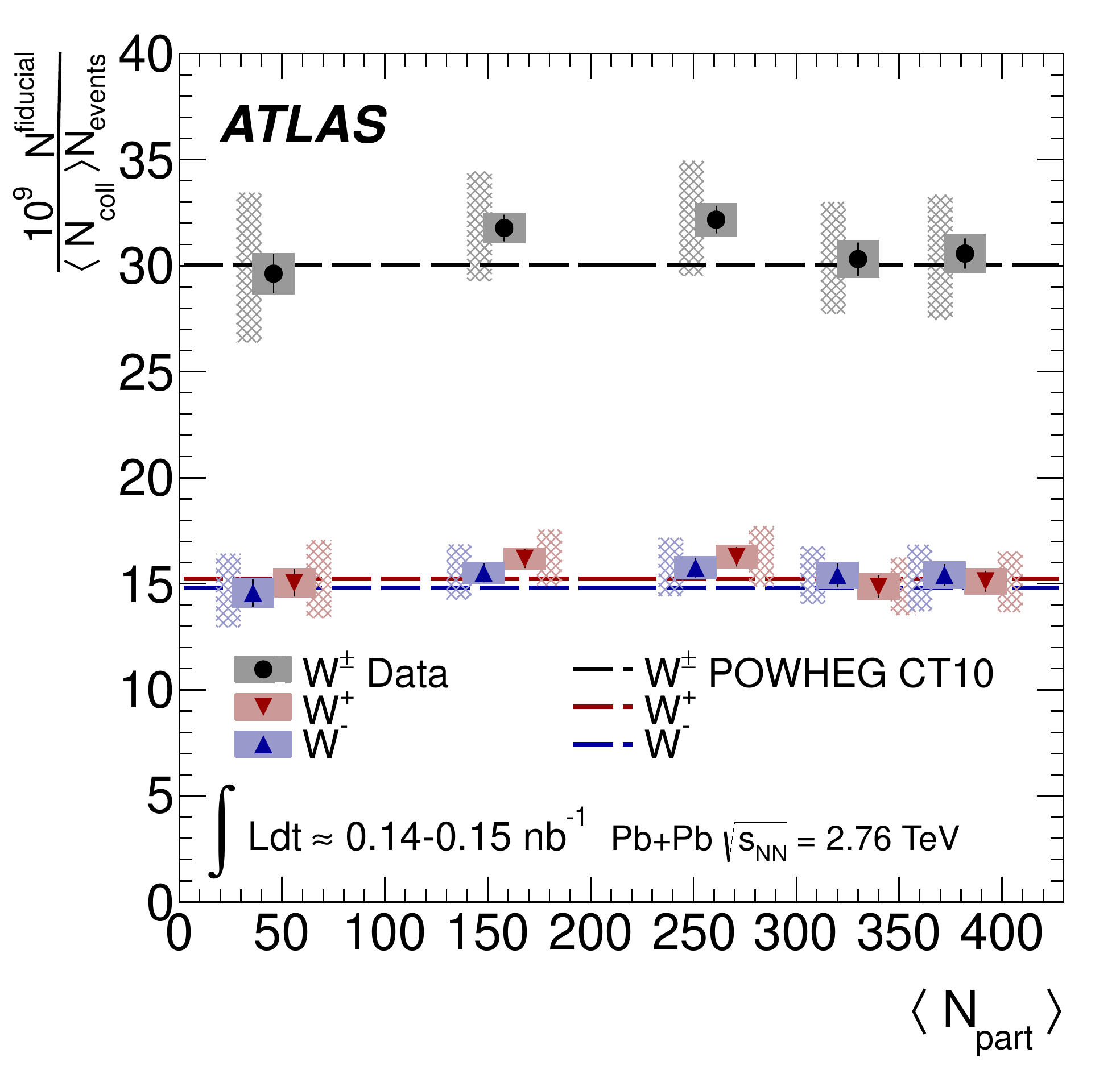}
\end{minipage}
\begin{minipage}{.29\textwidth}
\includegraphics[width=\textwidth]{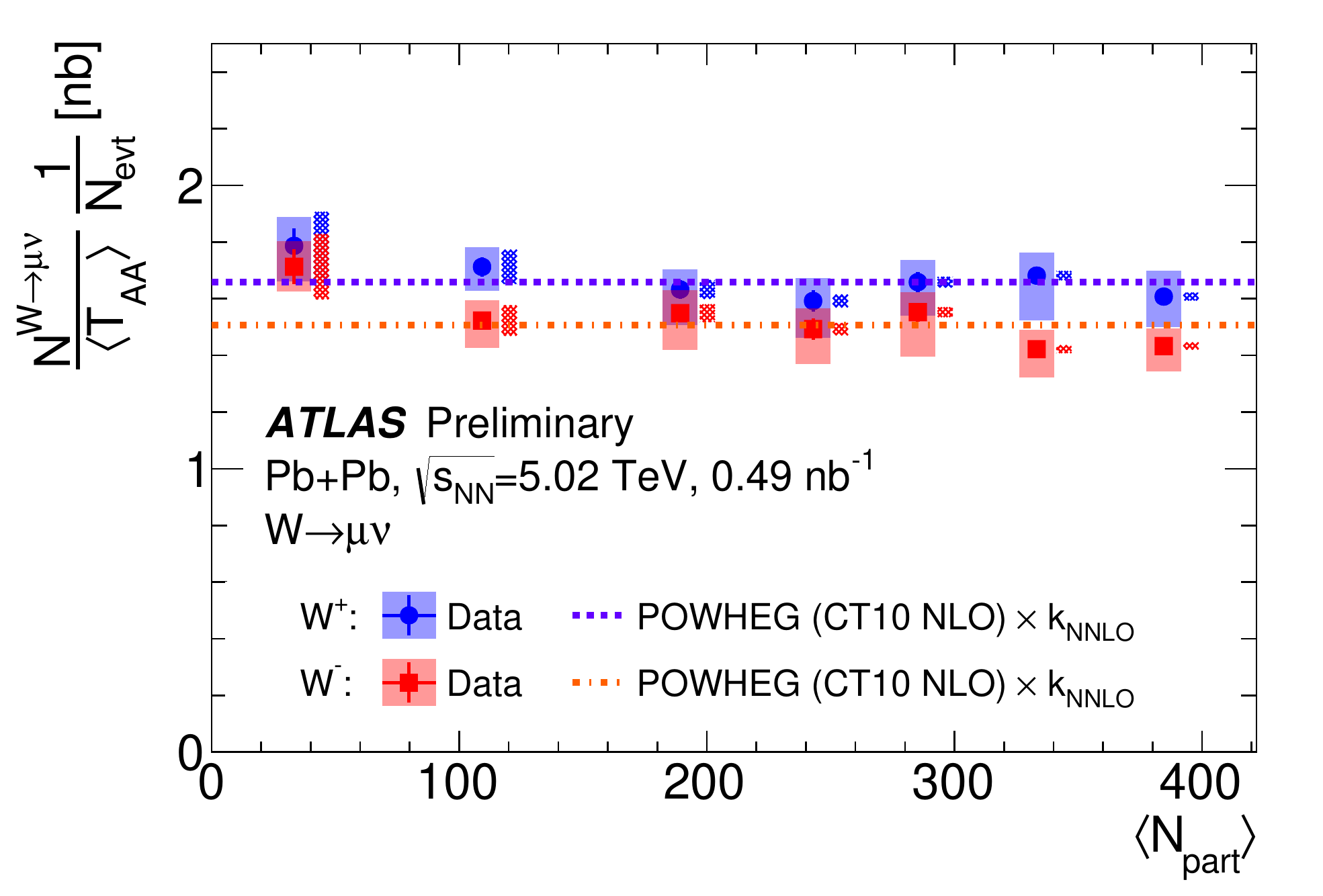}
\end{minipage}
\end{center}
\goup
\caption{Left panels: $Z$ boson yields in \PbPb divided by \Ncoll measured at two different \sqn. Right panels: analogues values for $W$ boson~\cite{PRL110-022301,ATLAS-CONF-2017-010}.
\label{fig:slide16}}
\end{figure}

A unique opportunity provided by EW bosons is the possibility to study the initial collision geometry. For the first time, the binary scaling in HI collisions was demonstrated with high accuracy using the $Z$ bosons at the LHC~\cite{PRL110-022301} and later confirmed with other measurements~\cite{ATLAS-CONF-2017-010}. This is shown with the $Z$ and $W$ bosons in Fig.~\ref{fig:slide16} 
as a ratio of yields divided by the number of binary collisions (\Ncoll) estimated using Glauber model~\cite{glauber}. The dependencies remain flat at all centralities at both measured \sqn. With increasing statistics of the HI samples it becomes possible to reduce both statistical and systematic uncertainties to a level that is smaller than the  uncertainties of the centrality determination. The results open opportunity to complement the nuclear modification factors based on Glauber model, with a model-independent double ratio of the production rates of any observable in the \NucNuc collision system to the \pp to the same ratio of the EW bosons.

\begin{figure}[htb!]
\begin{center}
\begin{minipage}{.24\textwidth}
\includegraphics[width=.9\textwidth]{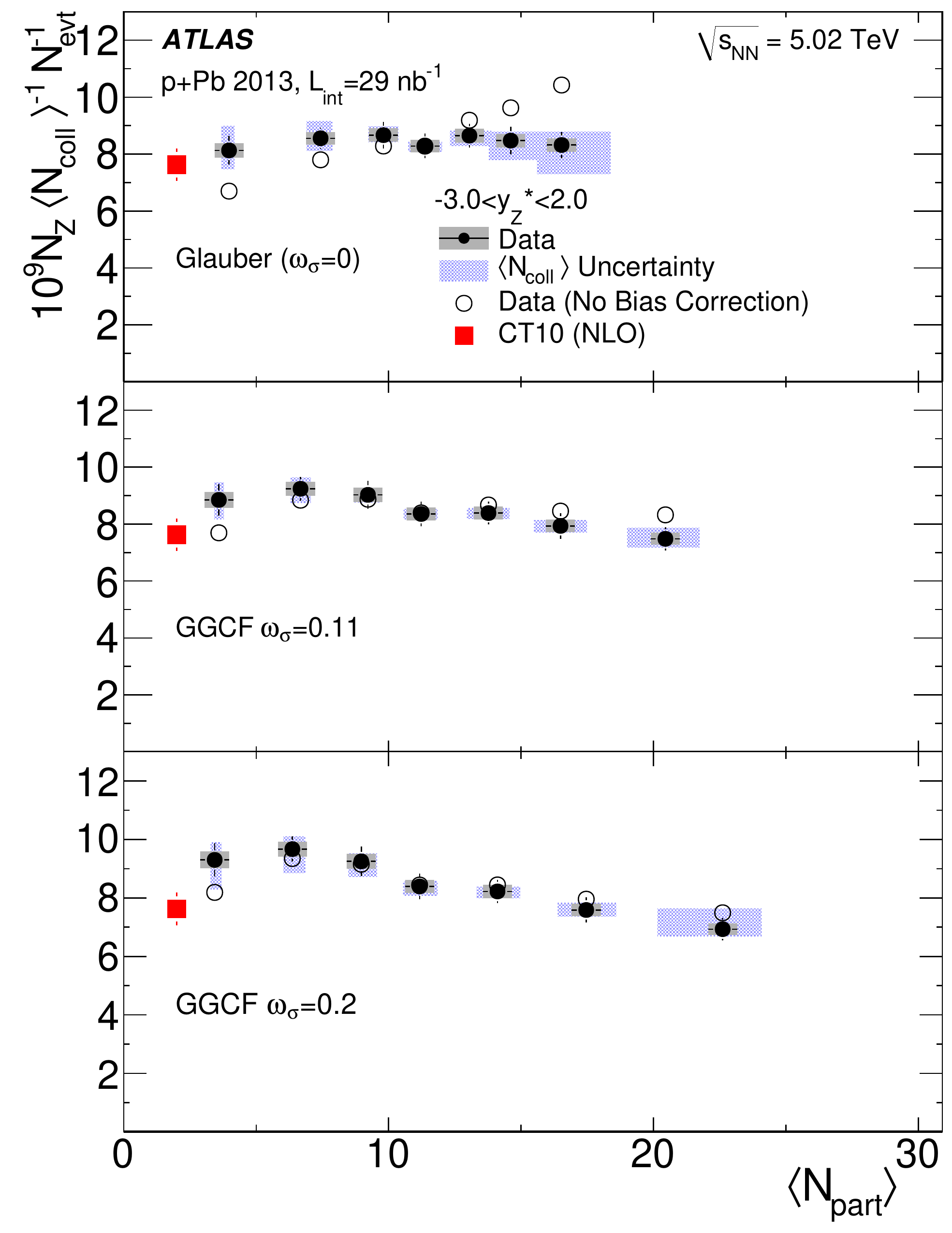}
\end{minipage}
\begin{minipage}{.24\textwidth}
\includegraphics[width=.9\textwidth]{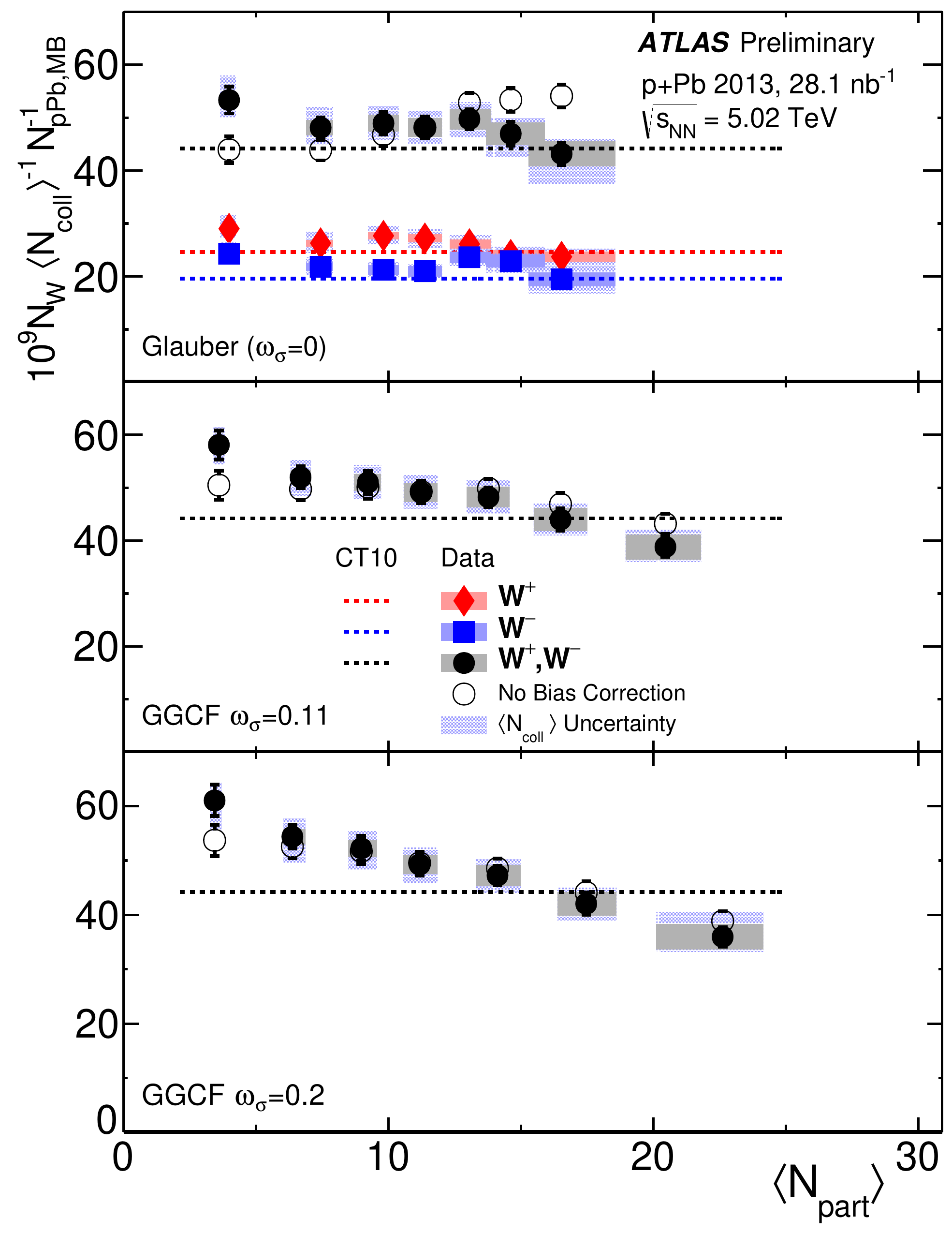}
\end{minipage}
\begin{minipage}{.44\textwidth}
\includegraphics[width=.9\textwidth]{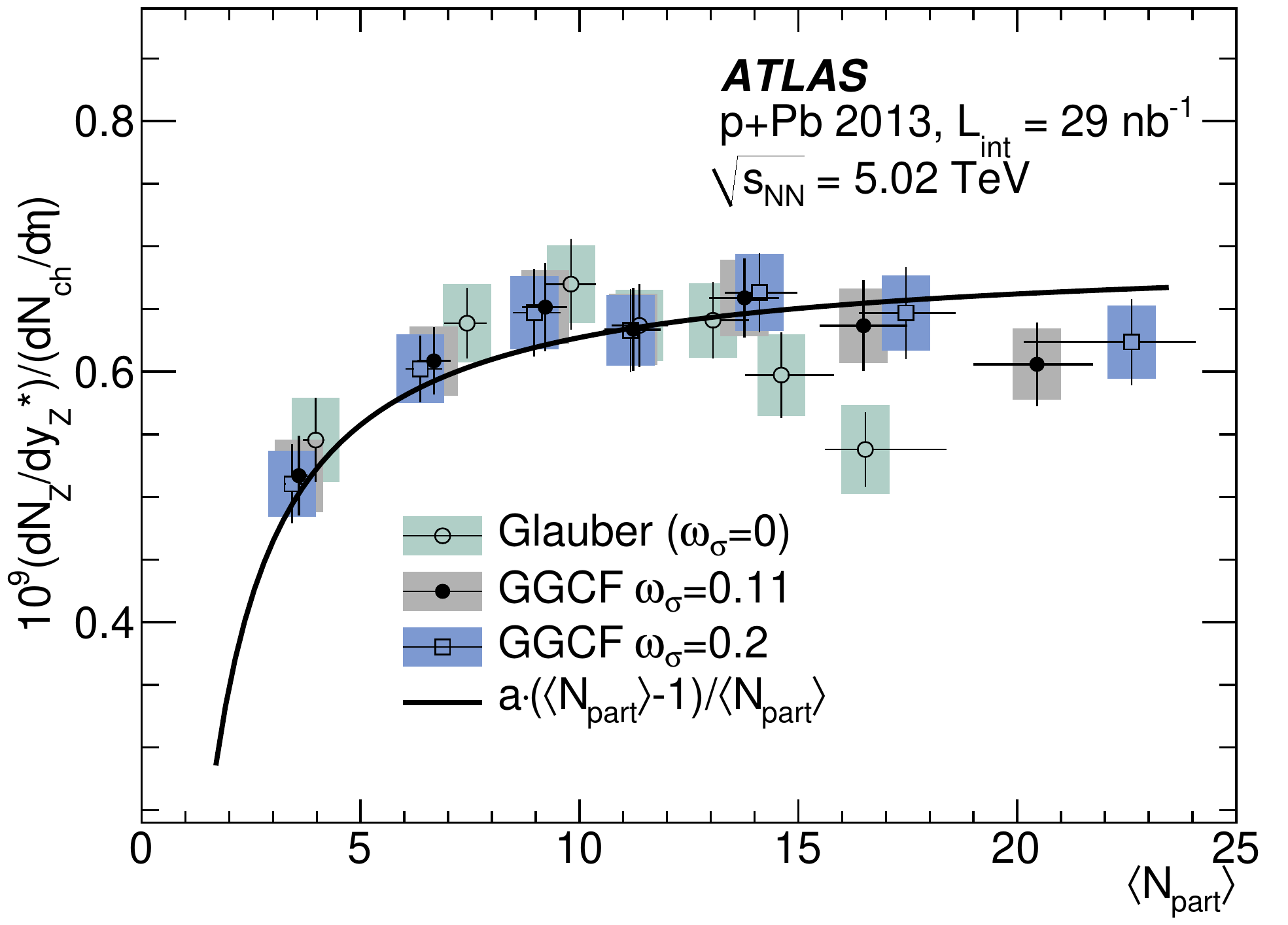}
\end{minipage}
\end{center}
\goup
\caption{Left two panels: $Z$ and $W$ boson yields in \pPb divided by \Ncoll. Right panel: $Z$-boson yield per charged particle~\cite{PRC92-044915,ATLAS-CONF-2015-056}.
\label{fig:slide17}}
\end{figure}

Interpretation of measured results on the accuracy to which the initial state geometry is understood is very evident in the \pPb data at $\sqn=5.02$~TeV. Shown in the left two panels of Fig.~\ref{fig:slide17} are the production rates of $Z$ and $W$ bosons as a function of centrality in \pPb collisions scaled by \Ncoll. 
The measurements are performed for the Glauber model (top panels) and for the two sets of parameters in the colour charge fluctuations models~\cite{ccfm1} (lower panels). Given present uncertainties, only the lower panel can be disfavoured by the data, but with the increased statistics of the 2016 \pPb data sample a more accurate conclusion can be made about the models. Data shown in the panels with open and closed markers are derived with and without the centrality bias correction which is based on Ref.~\cite{cbias}. Use of this correction is needed to get the correct shape of the curves, but it brings to the experimental measurement additional assumptions and uncertainties, which can be avoided in double ratios.

Another look at the geometric quantities in \pPb collisions is shown in the right panel of Fig.~\ref{fig:slide17}, where $Z$-boson production is compared with the rate of inclusive hadron production. According to the \Ncoll scaling, the ratio follows $(\Npart-1)/\Npart$, since in \pPb collisions proton contributes one count to $\Npart = \Ncoll -1$. The curve is consistent with the data.

\section{Impact parameter in \pp}
Special attention in recent years is paid to the small systems. Observation of the long-range correlations in \pp and \pPb systems suggests that the QGP-like effects can be formed in small systems. To prove or disprove this concept more experimental measurements are required. Also needed is the control over \pp system size that is currently done by the final state charged multiplicity. All existing data show weak or no dependence of the long-range correlation magnitude on multiplicity. Some theoretical predictions suggest that the initial state geometry is not necessarily correlated with the charged particle multiplicity and therefore a different approach is required to constrain the geometry~\cite{welsh}. Measurement done in ATLAS takes as an assumption that the impact parameter of the collision of events with registered $Z$ boson on average is smaller than in inclusive events. To obtain statistically significant $Z$--boson sample for measuring correlations, the entire $\sqs=8$~TeV \pp data with 19.6 fb$^{-1}$ is used. These data, taken at nominal LHC luminosity, contains significant pileup of about 20 interaction per bunch crossing.

\begin{figure}[htb!]
\begin{center}
\includegraphics[width=.6\textwidth]{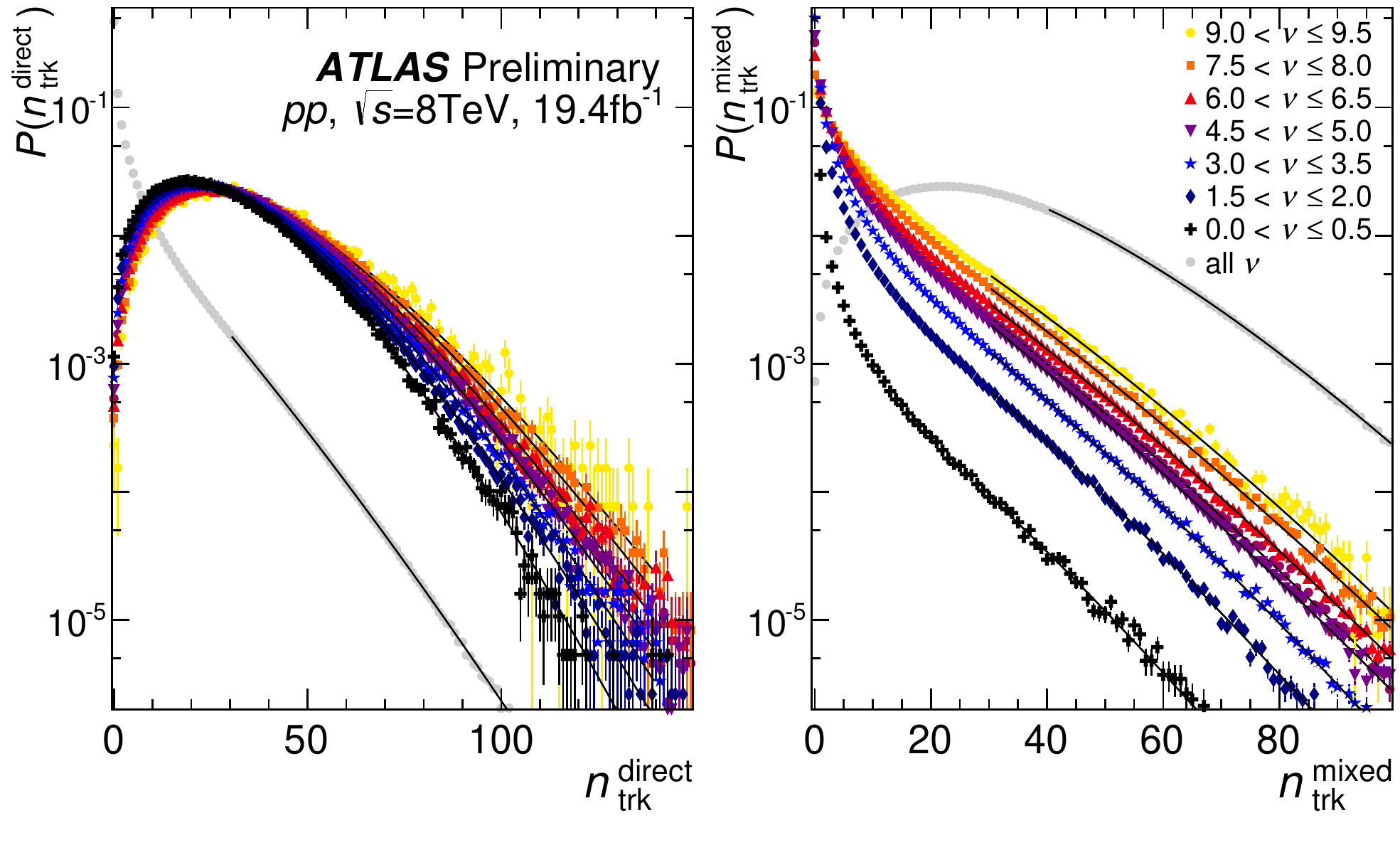}
\end{center}
\goup
\caption{Left: Distributions of the number of charged particle tracks measured in the data. Right: same for mixed events, equivalent to the pileup. Colours correspond to different average numbers of pileup tracks, grey color shows complementary inclusive distribution~\cite{ATLAS-CONF-2017-068}.
\label{fig:slide24}}
\end{figure}

\begin{figure}[htb!]
\begin{center}
\includegraphics[width=.75\textwidth]{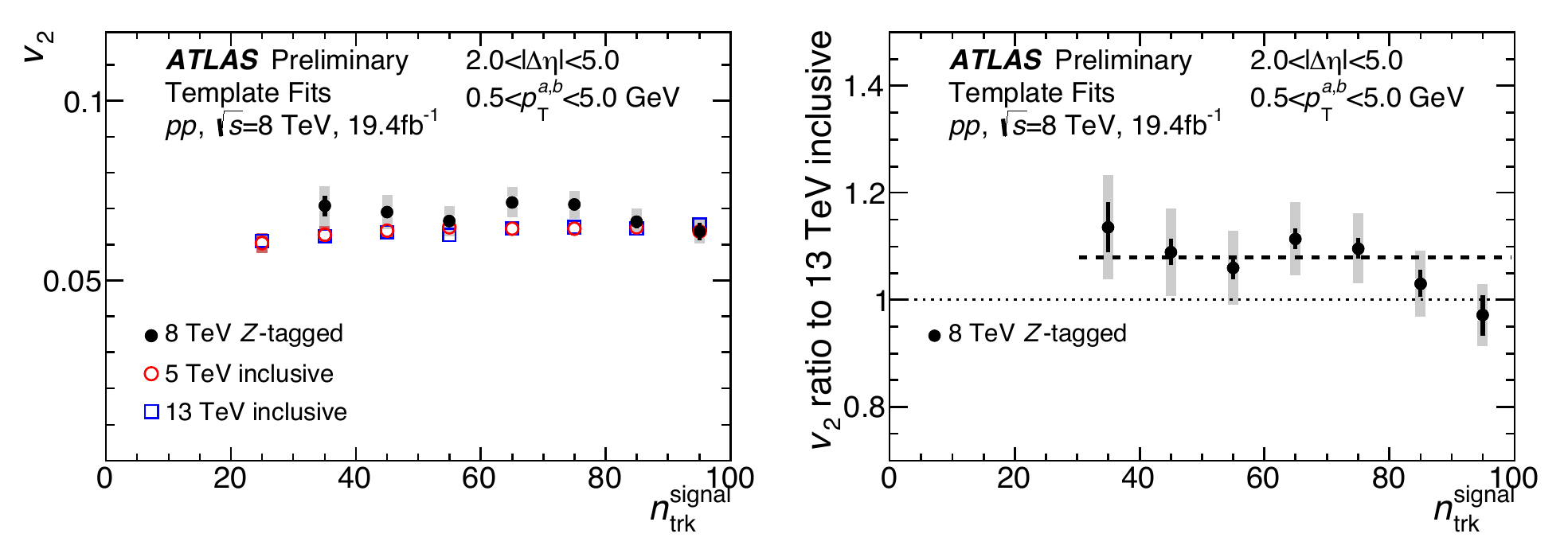}
\end{center}
\goup
\caption{Left: Distributions of the number of charged particle tracks measured in the data. Right: same for mixed events, equivalent to the pileup. Colours correspond to different average numbers of pileup tracks, grey color shows complementary inclusive distribution~\cite{ATLAS-CONF-2017-068}.
\label{fig:slide30}}
\end{figure}

The analysis uses novel procedure developed by ATLAS~\cite{ATLAS-CONF-2017-068}. The pileup is rejected by selecting only those particle whose track is compatible with the $Z$--boson vertex. The contribution of the residual pileup is evaluated and subtracted using mixed event technique. Unlike in many previous studies, in this analysis, a sample of the mixed event is produced such that mixed events have parameters distributions, such as multiplicity, kinematics, and correlations identical to the pileup. These events are constructed by selecting tracks found in data event $i$, in a vicinity of the $Z$--boson vertex position measured in data event $j$, where events $i\neq j$ are recorded at the same instantaneous luminosity. More procedure details are given in Ref.~\cite{ATLAS-CONF-2017-068}. This data-driven approach produces distributions shown in Fig.~\ref{fig:slide24}. 
Left and right panels show direct and mixed events respectively that are categorized by the parameter $\nu$, the average number of pileup tracks in the sample. Analysis of the two particle correlations is then done independently in each category of $\nu$ for different combinations of track pairs: two direct tracks, two mixed tracks from the same event, and also for pairs of tracks from direct and mixed events and from two different mixed events. The two latter contributions, by definition, contain no correlation, however proper accounting of all partial contributions allows deriving the final answer shown in Fig.~\ref{fig:slide30}. 
The ratio of the $v_{2}$ in events with $Z$ bosons to inclusive is $1.08\pm0.06$, which also is the first HI results measured with full LHC luminosity data.

This research is supported by the Israel Science Foundation (grant 1065/15), Israeli Committee for High Energy (grant IASAH-712760), and by the MINERVA Stiftung with the funds from the BMBF of the Federal Republic of Germany.

\end{document}